\documentclass[%
reprint,
 amsmath,amssymb,
 aps,
prapplied,
]{revtex4-2}

\usepackage{graphicx}

\usepackage{dcolumn}% Align table columns on decimal point
\usepackage{bm}% bold math
\begin{document}

\preprint{APS/123-QED}

%\preprint{APS/123-QED}

\title{Broad-band High-Energy Resolution Hard X-ray Spectroscopy \\using Transition Edge Sensors at SPring-8}% Force line breaks with \\

\author{Shinya Yamada}
\email{syamada@rikkyo.ac.jp}
\altaffiliation[]{Department of Physics, Tokyo Metropolitan University, Hachioji, Tokyo, 192-0397, Japan}
\author{Yuto Ichinohe}
\affiliation{Department of Physics, Rikkyo University, Toshima-Ku, Tokyo, 171-8501, Japan}

\author{Hideyuki Tatsuno}
\author{Ryota Hayakawa}
\author{Hirotaka Suda}
\author{Takaya Ohashi}
\author{Yoshitaka Ishisaki}
\affiliation{Department of Physics, Tokyo Metropolitan University, Hachioji, Tokyo, 192-0397, Japan}

\author{Tomoya Uruga}
\author{Oki Sekizawa}
\author{Kiyofumi Nitta}
\affiliation{Center for Synchrotron Radiation Research, Japan Synchrotron Radiation Research Institute (JASRI), Sayo, Hyogo, 679-5198, Japan}
\author{Yoshio Takahashi}
\author{Takaaki Itai}
\author{Hiroki Suga}
\author{Makoto Nagasawa}
\author{Masato Tanaka}
\author{Minako Kurisu}
\affiliation{Department of Earth and Planetary Sciences, Graduate School of Science, The University of Tokyo, Bunkyo-Ku, Tokyo, 113-0033, Japan}

\author{Tadashi Hashimoto}
\affiliation{Advanced Science Research Center (ASRC), Japan Atomic Energy Agency (JAEA), Tokai, Ibaraki, 319-1184, Japan}

\author{Douglas Bennett}
\author{Ed Denison}
\author{William (Randy) Doriese}
\author{Malcolm Durkin}
\author{Joseph Fowler}
\author{Galen O'Neil}
\author{Kelsey Morgan}
\author{Dan Schmidt}
\author{Daniel Swetz}
\author{Joel Ullom}
\author{Leila Vale}
\affiliation{Quantum Sensors Group, National Institute of Standards and Technology (NIST), Boulder, Colorado, 80305, USA}

\author{Shinji Okada}
\affiliation{Engineering Science Laboratory, Chubu University, Kasugai, Aichi, 487-8501, Japan}
\altaffiliation[]{Cluster for Pioneering Research, RIKEN, Wako, Saitama, 351-0198, Japan}

\author{Takuma Okumura}
\author{Toshiyuki Azuma}
\author{Toru Tamagawa}
\altaffiliation[]{Nishina Center, RIKEN, Wako, Saitama 351-0198, Japan}
\affiliation{Cluster for Pioneering Research, RIKEN, Wako, Saitama 351-0198, Japan}

\author{Tadaaki Isobe}
\affiliation{Nishina Center, RIKEN, Wako, Saitama, 351-0198, Japan}

\author{Satoshi Kohjiro}
\affiliation{Superconducting Sensors and Circuits Group, National Institute of Advanced Industrial Science and Technology (AIST), Tsukuba, Ibaraki, 305-8568, Japan}

\author{Hirofumi Noda}
\affiliation{Department of Earth and Space Science, Osaka University, Toyonaka, Osaka, 560-0043, Japan}
\author{Keigo Tanaka, Akimichi Taguchi}
\affiliation{College of Science and Engineering, Kanazawa University, Kakuma-machi, Kanazawa, 920-1192, Japan}
\author{Yuki Imai, Kosuke Sato}
\affiliation{Department of Physics, Saitama University, Saitama-shi, Saitama, 338-8570, Japan}
\author{Tasuku Hayashi}
\affiliation{Astromaterials Science Research Group (ASRG), Institute of Space and Astronautical Science (ISAS), Japan Aerospace Exploration Agency (JAXA), Sagamihara, Kanagawa, 252-5210, Japan}
\author{Teruhiko Kashiwabara}
\affiliation{Submarine Resource Research Center, Research Institute for Marine Resources Utilization, Japan Agency for Marine-Earth Science and Technology (JAMSTEC), Yokosuka, Kanagawa, 237-0061, Japan}
\author{Kohei Sakata}
\affiliation{Center for Global Environmental Research, National Institute for Environmental Studies (NIES), Tsukuba, Ibaraki, 305-8506, Japan}

\date{\today}% It is always \today, today,

\begin{abstract}
We have succeeded in operating a transition-edge sensor (TES) spectrometer 
and evaluating its performance at the SPring-8 synchrotron X-ray light source. 
The TES spectrometer consists of a 240 pixel National Institute of Standards and Technology (NIST) TES system, 
and 220 pixels are operated simultaneously with an energy resolution of $4$~eV at 6~keV at a rate of $\sim$1~c/s/pixel. 
The tolerance for high count rates is evaluated in terms of energy resolution and live time fraction, 
leading to an empirical compromise of $\sim2\times10^3$~c/s/all pixels 
with an energy resolution of $5$~eV at 6~keV. 
By utilizing the TES's wide-band spectroscopic capability, simultaneous multi-element analysis is demonstrated for a standard sample. 
We conducted X-ray absorption near-edge structure (XANES) analysis in fluorescence mode using the TES spectrometer.
The excellent energy resolution of the TES enabled us to detect 
weak fluorescence lines from dilute samples and trace elements that have previously been 
difficult to resolve due to the nearly overlapping emission lines of other dominant elements. 
The neighboring lines of As K$\alpha$ and Pb L$\alpha2$ of the standard sample were clearly resolved and 
the XANES of Pb L${\alpha}2$ was obtained. 
Moreover, the X-ray spectrum from the small amount of Fe in aerosols was 
distinguished from the spectrum of a blank target, which helps us to understand the targets and the environment. 
These results are the first important step for the application of high resolution TES-based spectroscopy at hard X-ray synchrotron facilities. 

\end{abstract}

\maketitle

\section{\label{sec:level1}Introduction}

X-ray absorption (XAS) and X-ray fluorescence (XRF) or X-ray emission spectroscopy (XES) are powerful techniques to measure distribution, 
chemical state, and local structure of elements. Therefore, they play essential roles in various materials analyses. 
For example, diagnostics for trace elements are conducted by measuring the energy spectrum of fluorescent X-rays. 
Instruments for performing X-ray spectroscopy are primarily categorized into wavelength-dispersive or energy-dispersive spectrometers. 
XAS of trace elements has typically been conducted using Si or Ge-based energy-dispersive X-ray detectors. 
The typical energy resolution of these detectors is about 120 eV in the hard X-ray energy region ($>$4 keV), 
which is fundamentally limited by statistical variation of the number of electron-hole pairs created by X-ray photon absorption.  
However, that resolution is insufficient to fully distinguish the nearly overlapping K and L emission lines of elements in this energy regions ($<$15 keV), 
which hinders the ability of trace element analysis to effectively measure natural or otherwise complicated samples. 
In cases where it is necessary to resolve closely-spaced emission lines, 
a Bent Crystal Laue Analyzer (BCLA) is used in combination with energy-dispersive X-ray detectors\cite{Takahashi2006}. 
The energy resolution of this system is several tens of eV. 
However, XES measurements require high energy resolution, less than 10 eV, 
and are therefore normally conducted using a wavelength-dispersive bent multi-crystal analyzer spectrometer. 
The detection efficiency of the BCLA was limited due to narrow acceptance angle and low diffraction efficiency of the analyzer crystals.
The analyzer crystals need to be changed depending on the energy region of interest. 
Recently, multi-crystal spectrometers based on intersecting Rowland circles have been developed and enhanced the solid angle and the detection efficiency (e.g., \cite{doi:10.1063/1.4803669} and \cite{Huotari:vv5158}). In such a system, the detectors and the crystals need to be located at a certain distance from the sample, 
which have a limit to increase the efficiency and the bandpass of the energy range. 

These situations demand high-efficiency energy-dispersive X-ray detectors having energy resolution less than 10 eV. 
The first technology to deliver this type of performance came in the form of semiconductor type microcalorimeters operating at 50~mK. 
To achieve high energy resolution, operating the detectors at such low temperature is essential 
since the creation energy of the pairs (e.g., $\sim$ eV for electron and hole pairs in semiconductors, and $\sim$ meV for Cooper pairs in superconductors) and the thermal noise decreases. 
Recently, the semiconductor type microcalorimeters, such as the Soft X-ray Spectrometer (SXS) on the X-ray astronomical satellite called ASTRO-H (Hitomi)\cite{10.1117/12.2055681}, 
succeeded in solving several issues related to vibration from the cryogenic system\cite{10.1117/1.JATIS.4.1.011216} and achieved an energy resolution of $4.5$~eV at 6~keV in a space environment. 
Its high energy resolution clearly resolved ionized iron lines and 
revealed that the hot plasma stored in the galaxy cluster has low turbulent pressure\cite{Hitomi2016}. 
Although there is now a great deal of accumulated experience using semi-conductor X-ray microcalorimeters, 
their high impedance limits the increase of the number of pixels for practical use, 
and hence the collection area is fundamentally limited 
because the heat capacity, and therefore the size, of a single pixel is also limited by its required energy resolution. 

A promising alternative candidate is the superconducting transition-edge sensor (TES)\cite{Irwin2005}. 
Since a way of stabilizing the TES within its narrow transition region was formulated in the mid-1990s \cite{doi:10.1063/1.117630}, 
the technical details and physical nature of these devices have been intensively studied, with strong demand from both on-ground and space applications\cite{Randy2017}. 
Thanks to these advances, the TES has been chosen as the detector technology for the future European X-ray observatory called ATHENA\cite{athena2015}, which is planned to have an array of more than three thousand sensors. 
However, in order to successfully integrate a TES array into other systems in an experiment, 
the spectrometer design must overcome several difficulties, including suppressing mechanical vibration, 
isolating electrical interference, shielding against magnetic fields, and synchronizing its timing with other systems. 
The National Institute of Standards and Technology (NIST) has developed a mature 240-pixel TES spectrometer, which was applied to nuclear physics research at Japan Proton Accelerator Research Complex (J-PARC)\cite{Okada2016} 
and succeeded in measuring fluorescence lines from a Kaonic atom\cite{7801889}, achieving stable performance even in the challenging environment found at an accelerator. 
The TES can be applied to a diverse area of scientific problems by changing its design, e.g., thickness or size of TESs and absorbers, to cover a wide range of energies and celestial or terrestrial X-ray sources. 
For example, the TES can be applied as a particle detector to search for a dark matter via inelastic scattering\cite{Maisonobe}, 
or to measure the mass of neutrinos\cite{Nucciotti}. 

Today, TES X-ray spectrometers are operated for several different applications: e.g., 
at the Stanford Synchrotron Radiation Lightsource beamline 10-1 (BL 10-1)\cite{doi:10.1063/1.5119155}, 
at the Electron Beam Ion Trap (EBIT) at NIST\cite{doi:10.1063/1.5116717}, 
for resonant soft X-ray scattering at the Advanced Photon Source\cite{PhysRevApplied.13.034026}, for particle-induced X-ray emission (PIXE)\cite{PhysRevApplied.6.024002}, 
and for ultrafast time-resolved X-Ray emission spectroscopy on a tabletop\cite{PhysRevX.6.031047}. 
However, compared to these applications, which are either targeted at soft X-ray science or  tabletop experiments in relatively small laboratories, the potential of TESs for applications at hard X-ray synchrotron facilities has not been explored. 
Although there were successful reports of using a crystal analyzer system to measure X-ray absorption fine structure (XAFS) in the hard X-ray, 
it was not ideal for several reasons: 
crystals were designed for each energy of emission lines, 
and needed precise adjustment each time, the available energy resolution was moderate, and the measurable energy range was narrow\cite{Takahashi2006}. 
TESs can contribute to the reduction of the alignment time, improve energy resolution, and expand the energy coverage. 
Note that though both technologies will progress, the advantage in using TES is that it does not require reflection at the crystals and hence relaxes spatial constraints. 
Here we will present the commissioning of a new TES spectrometer at the beamline BL37XU at SPring-8, Japan\cite{doi:10.1063/1.1757812}. 
The light source of BL37XU can provide hard X-ray photons with an energy range of 4.5-18.8~keV by using 1st harmonic X-ray of undulator, 
which is suitable to measure a sample via K$\alpha$ lines from relatively heavy metals and L-lines from rare earth elements. 
Fluorescence-mode X-ray Absorption Near-Edge Structure (XANES) of dilute elements in various samples was difficult to measure using conventional detectors such as Silicon Drift Detector (SDD)
Here we obtain the XANES signal of a dilute aerosol sample that did not work well with the SDD detector.
In this paper, we present in detail the set-up and data processing procedure, 
performance of each pixel of the TES, and its application to the simultaneous multi element analysis of XANES of heavy elements and dilute samples. 

\begin{figure*}[htbp]
\begin{center}
\includegraphics[width=0.98\linewidth]{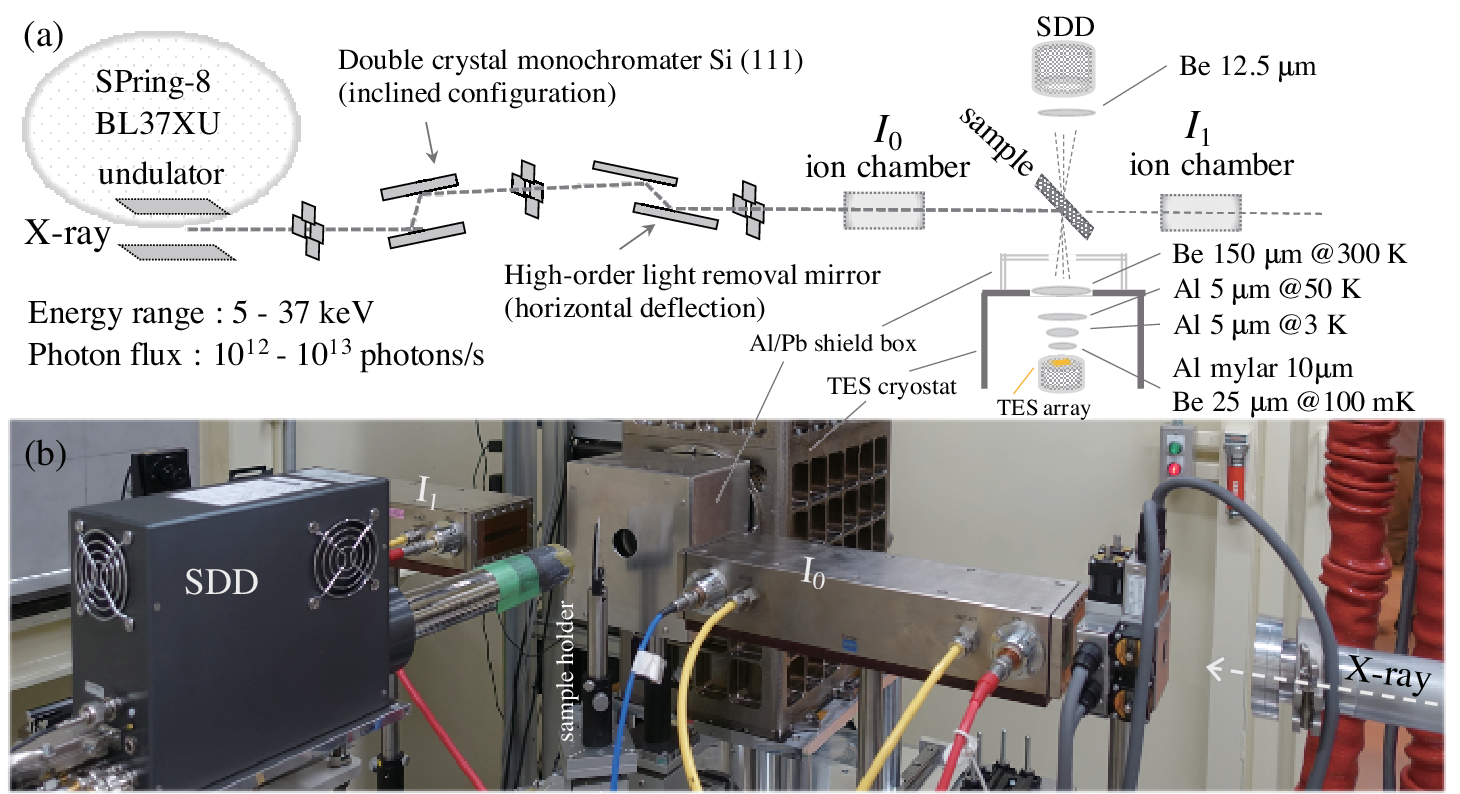}
\caption{(a) Overview of the experimental setup at BL37XU at SPring-8. The TES is set at the front side of the sample at an angle of 45 degrees, while 
the SDD is on the opposite side. (b) The picture of the setup. The TES cryostat is covered with an aluminum box with a lead shield inside to prevent stray light from reaching the detector.} 
\label{config}
\end{center}
\end{figure*}

\section{\label{sec:level2}Experiments}
\subsection{\label{sec:level2-1}Detector setup}

A quasi-monochromatic X-ray beam emitted from the SPring-8 standard undulator (period=3.2~cm, $B_{\rm{max}}=0.78$~Tesla) was monochromatized to a certain energy (5--37 keV) using a Si(111) monochromator. 
BL37XU is a hard X-ray undulator beamline in Spring-8 that is mainly used for studies of X-ray micro/nano-spectrochemical analysis 
such as XRF/XAFS imaging, wavelength-dispersive XAFS, depth-resolved XAFS, and XRF holography. 
The flow of the incoming X-rays in the BL32XU is shown in figure \ref{config} (a). 
Two horizontal deflecting mirrors are placed downstream of the monochromator in order to 
eliminate higher harmonics and to obtain a focused X-ray beam in horizontal direction. 
There are several slits with mechanically movable remote-motors, which can reduce the intensity of the beam. 
The relative energy resolution of the incident X-ray beam, $\Delta E/E$, is approximately $2\times10^{-4}$, which is proportional to $\sqrt{\sigma_{\rm{crystal}}^2+\sigma_{\rm{beam}}^2}$,  
where $\sigma_{\rm{crystal}}$ is Darwin width of monochromator crystal (e.g., 20--60 $\mu$rad @14--5 keV), and $\sigma_{\rm{beam}}$ is an angular divergence of incident X-ray beam (e.g., 5 $\mu$rad with a slit width of 200 $\mu$m and 40 m away from the source). Thus, the effect of operating the slits on the energy resolution of the incident X-ray beam could be at most 3\%, $\sim$0.04 eV@6keV. 
Samples and the X-ray windows on sample side used for both the SDD and TES spectrometer are exposed to ambient air at room temperature. 

Since a TES makes use of a thermal process as a measure of photon energy, 
the speed of the relaxation of the heat deposited in the X-ray absorber limits the response time. 
In general, this type of detector is difficult to tune to work as fast as a SDD that works using electromagnetic interaction. 
We used several slits along the beamline to suppress the beam intensity. 
The vacuum jacket of the cryostat or the Dewar, whose size is 33cm$\times$22cm$\times$66cm, is mounted on a movable support structure. 
This flexibility enables us to tune the X-ray intensity by two orders of magnitude. 
The setup is presented in figure \ref{config}. 
The sample was set at 45 degrees with respect to the incident X-ray. 
The fluorescence that escapes from the surface of the sample is transmitted to the TES array.
Since SDD has a larger effective area, it is positioned to receive photons from the backside of the sample.
Two ion chambers are used for the measurement of intensities of an incident X-ray ($I_0$) and a transmitted X-ray ($I_1$). 

Each TES consists of a superconducting bilayer of thin Mo and Cu films. 
An X-ray absorber of 4~$\mu$m thick Bi is deposited on top of the TES. 
Each pixel has an active area of 320~$\mu$m$\times$305~$\mu$m, 
which is determined by a gold-coated 275~$\mu$-thick Si aperture chip placed on top of the TES array, 
which prevent X-rays from hitting the array outside of the X-ray absorbers; 
otherwise, X-ray heating of the substrate could deteriorate the energy resolution. 
The holes of the aperture chip are patterned by dry etching to have the same shape as the TES absorbers, 
so it does not obscure the X-rays passing from the holes to the TES array. 
The total active area of the 240-pixel array is about 23.42~mm$^2$. 

The detectors are cooled through a combination of a pulse tube cooler (Cryomech PT407 with a remote motor) 
and an adiabatic demagnetization refrigerator (ADR). 
Cooling from room temperature takes 17 hours, plus an additional one hour to cycle the ADR. 
This results in a TES bath temperature of 75 mK $\pm$ 4 $\mu$Krms. 
The ADR also provides an additional cold stage at 500~mK, where some of the readout electronics are mounted. 
The TES pixels are then electrically biased to their superconducting critical temperature of $T_C \sim$ 100~mK.

Since the normal resistance of the TES is on the order of $\sim$ 10 m$\Omega$ 
and its change caused by the X-ray is on the order of $\sim$ m$\Omega$, 
the signal needs to be read out by a low-impedance amplifier, such as a SQUID.
The voltage fed back to the feedback coil coupled to the SQUID is proportional to the time derivative of the resistance initiated by absorption of an X-ray. 
Details of the room temperature electronics are summarized in the reference\cite{2003RScI744500R}. 
The TES array uses a time-division-multiplexing readout system, which samples the current signal of the 240 sensors by dividing them into 8 SQUID columns.
The sampling time of each sensor is 7.2~$\mu$s (=240~ns $\times$ 30 pixels), and thus the effective sampling rate is 139 kHz.
A record of 1024 samples (= 7.3728 ms) was captured for each X-ray event. This length is changeable to any value in the software so that it can be chosen 
to match the length of the X-ray pulses.
Unlike with a SDD, 
to obtain optimal energy resolution from a TES requires applying a matched filter 
(optimal filter)\cite{1993JLTP93281S}\cite{Fowler2016} to the detector signal.
Currently, the room-temperature hardware writes the entirety of these records for every trigger event to a storage disk at high speed, and off-line software is used to do post-processing of pulse-shape analysis to achieve the best possible energy resolution.  
This implementation of saving the X-ray pulses causes the data quantity produced by the TES to be larger by three orders of magnitude than that of a SDD. 
 
\begin{figure}[htbp]
\begin{center}
\includegraphics[width=0.99\linewidth, keepaspectratio]{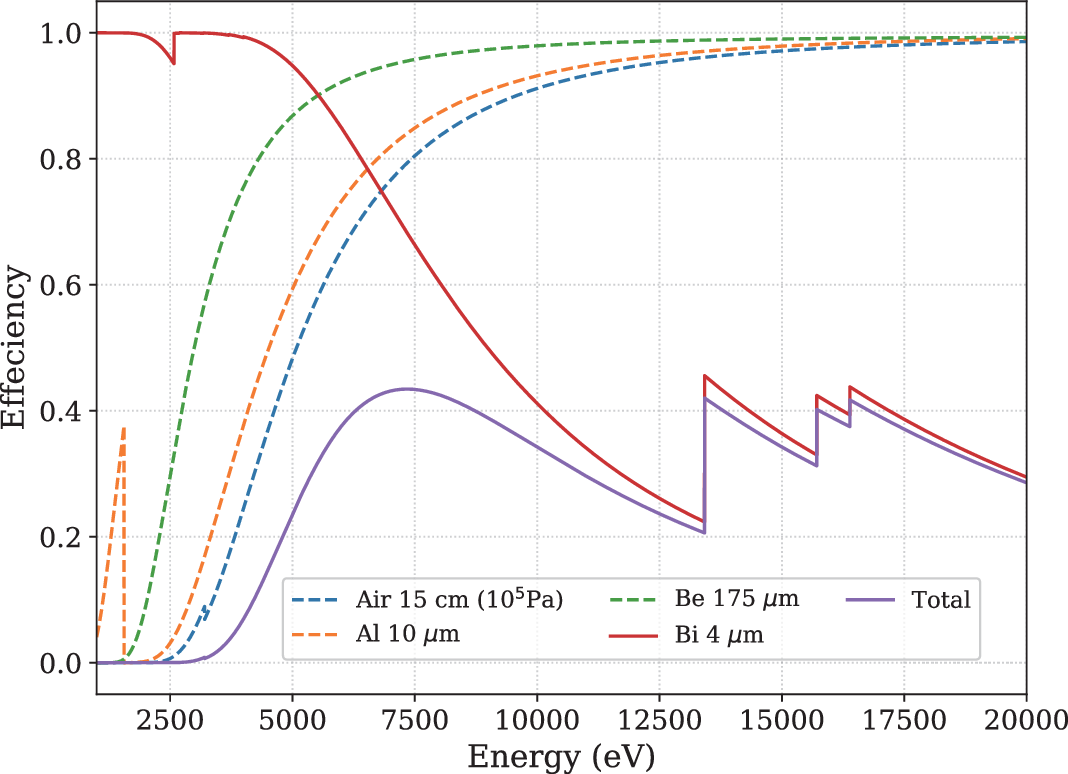}
\caption{The theoretical quantum efficiency of the TES is shown. The transmission of Air (15~cm, 10$^5$~Pa), 
Al filters (5~$\mu$m @ 3~K and 5~$\mu$m @ 60~K), 
Be windows (150~$\mu$m @300~K and 25~$\mu$m @100~mK), 
the photo-absorption efficiency of 4~$\mu$m-thick Bi absorber, and the total efficiency are shown in blue dotted, yellow dotted, green dotted, red solid, and purple solid lines, respectively.} 
\label{tesqe}
\end{center}
\end{figure}

The X-ray absorber and TES are cooled at the coldest stage of the ADR. There are four windows 
between room temperature and the cold stage that are intended to reduce IR/vis/UV light on the TESs 
while transmitting most X-rays; a 10$\mu$m aluminized mylar and 25$\mu$m Be at 50~mK, 5$\mu$m Al at 3~K, 5$\mu$m Al at 50~K, 
and a 150~$\mu$m Be window at the vacuum shield at room temperature. 
The detection efficiency is shown in figure \ref{tesqe}. 
The 50~mK shell is made of Al superconducting magnetic shield except for the X-ray window, 
though Al in the aluminized mylar might work to some extent.  
The distance of the TES to the sample is 15 cm. 
In this setup, the air outside the evacuated cryostat absorbs X-rays more than the filters. 
The quantum efficiency of $4\mu$-thick Bi is $\sim$80 \% at 6 keV and $\sim$20\% at 13 keV, 
while it increases to $\sim$40\% above the L-edges of Bi.  
There are two limiting factors on the largest detectable energy; 
the temperature of TES increases above $T_C$ due to huge thermal input from X-ray 
and the input signal to SQUID exceeds its linear range. 
The limit also depends on the bias voltage applied to the TES. 
We set the bias voltages to keep $\sim$20\% of the normal resistance of TES at $T_C$. 
Using the setup, photons below about 18 keV can be measured with the system. 

The TES, which is a thin film in an intermediate phase between normal and superconducting state, is sensitive to small magnetic fields, 
so a few percent of the Earth’s magnetic field could deteriorate the sharp edge of the transition.
There are two magnetic shields made of high permeability metal at the 50~K and 3~K shells, 
and a shield made out of a Type I superconductor at 50~mK. 
The shielding effect is locally weak at the apertures for X-rays  
since they are made of thin filters. 
Therefore, if there is a leak of a magnetic field along the incoming X-ray path, 
it could worsen the performance of the detector. 
However, at the BL37XU at SPring-8, the detector worked without any empirical sign of magnetic interference through the apertures.

\subsection{\label{sec:level2-1}Energy resolution}

The limit of the energy resolution of a TES is proportional to the temperature of the detector and the square root of heat capacity, 
while the total energy range is positively correlated with the heat capacity.
The thickness of the X-ray absorber and the size is designed to meet the scientific requirement for a particular experiment. 
The TES used in this experiment was optimized for the detection of 6 keV X-rays. 
In an actual measurement, the energy resolution does not reach the theoretically derived value. 
This is because, as is typical in other low-temperature detectors, 
electrical and magnetic interference, or mechanical micro-vibration can increase additional noise terms. 
This is one of the technical challenges of the application of the TES. 
Reproducing the best detector performance when the spectrometer is moved from one to another place 
requires a detailed design of the TES system and the method for integrating it into the facility. 

\begin{figure}[htbp]
\begin{center}
\includegraphics[width=0.99\linewidth]{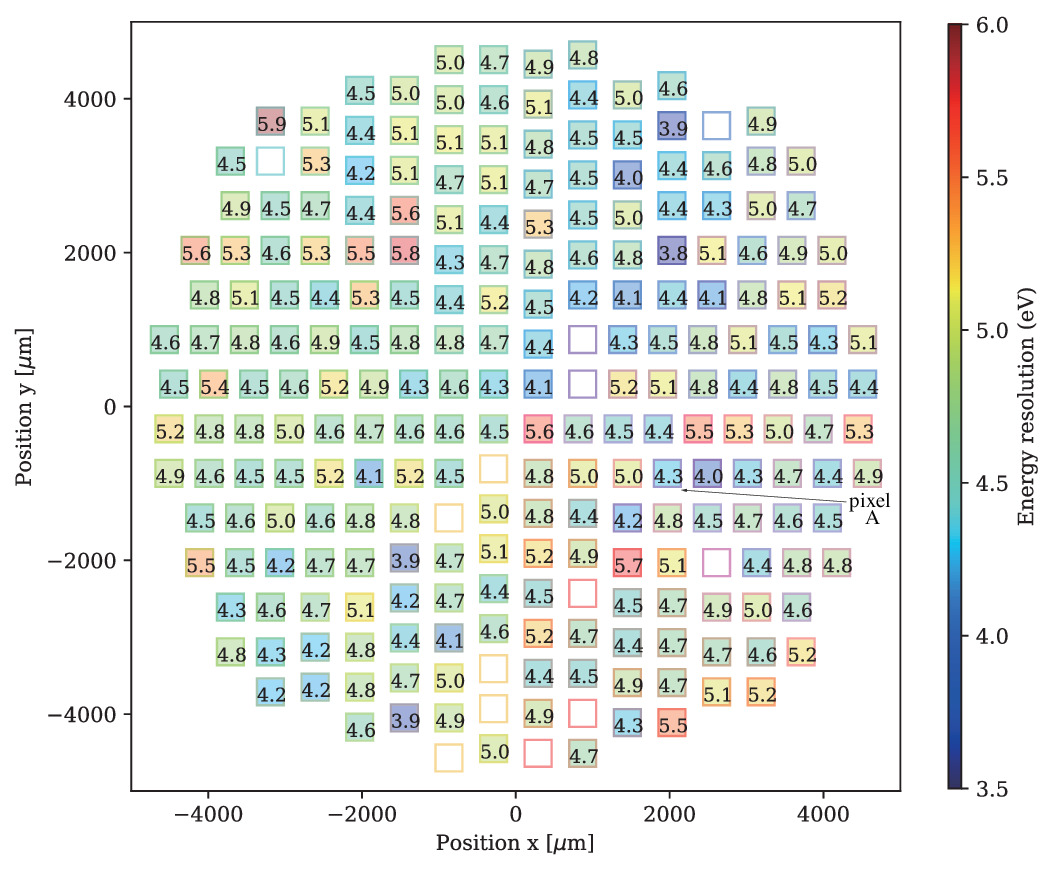}
\caption{The map of the energy resolution of the TES pixels. This map reflects the physical location of each pixel, with each sensor's energy resolution is shown on its position. The color bar linearly scales with the energy resolution. The pixels which were not operated are not colored. The location of pixel A is indicated with the arrow. } 
\label{resol}
\end{center}
\end{figure}

In our case, based on the experiment led by the High-resolution Exotic Atom X-ray spectroscopy with Transition-Edge Sensors (HEATES) collaboration\cite{Yamada2020}, 
the electrical interface of the vacuum chamber and the detector system was designed to be electrically isolated from both the pulse tube's compressor and the equipment of the beamline when it was installed. 
When we confirmed the basic performance in our laboratory, the compressor was electrically isolated from the cryostat. 
However, in practice, this configuration may not be the best due to the different grounding conditions. 
In SPring-8, we found that the temperature fluctuation of the 75 mK stage was 4~$\mu$Krms when the TES was grounded to the ground of the beamline, 
whereas the fluctuations increased by an order of 2 when floated. 
The isolation transmitters which transfer electric surges by magnetic coupling and the power stabilizers which are capable of suppressing high frequency and regulating voltage automatically were inserted into the power lines of the TES system. 
During commissioning, this configuration gave temperature fluctuations of $\sim4~\mu$Krms and no additional noise components in the detectors. 
Note that this configuration means that the compressor of the pulse tube, the SDD, and other electronics in the beamlines were connected to TES through low impedance, thus they could affect the performance of the TES if any of them are not properly grounded. 

The energy resolution is evaluated using $^{55}$Fe isotopes in this configuration.    
Although the K${\alpha1}$ and K${\alpha2}$ of Mn can be modeled with two lines when it is in a neutral gaseous state, as a solid the emission spectrum must be fitted with a model consisting of a sum of eight Voigt functions which is empirically derived by crystal spectrometers\cite{1997PhRvA564554H}. 
The TES is assumed to have a Gaussian response. However, the Bi absorber is known to have a low energy tail, so the tail effect is taken into consideration according to Tatsuno et al. 2016\cite{Tatsuno2016}. 
The low energy tail is caused by trapping of some of the heat carriers along grain boundaries in the evaporated Bi absorber.  
Recently, this issue has shown to be solved by using bismuth electroplated onto a Au seed layer, which increases the bismuth grain size, which nearly eliminates the low energy tail\cite{doi:10.1063/1.5001198}. The typical amount of the tail in our TESs is $\sim$20\% at 6 keV, and the exponential decay length is $\sim$ 20~eV. 
The energy resolution of each pixel is measured by including Gaussian broadening and the low energy tail in the response function, 
and presented as a physical pattern of the detector in figure \ref{resol}. 
The energy resolution is expressed as full width at half maximum (FWHM) throughout this paper. 
Some of the pixels were not electrically connected or not biased due to physical or electrical problems. 
These energy resolutions are equivalent to the best values obtained by this instrument at the J-PARC hadron environment.

% energy vs. dE
\begin{figure}[htbp]
\begin{center}
\includegraphics[width=0.98\linewidth, keepaspectratio]{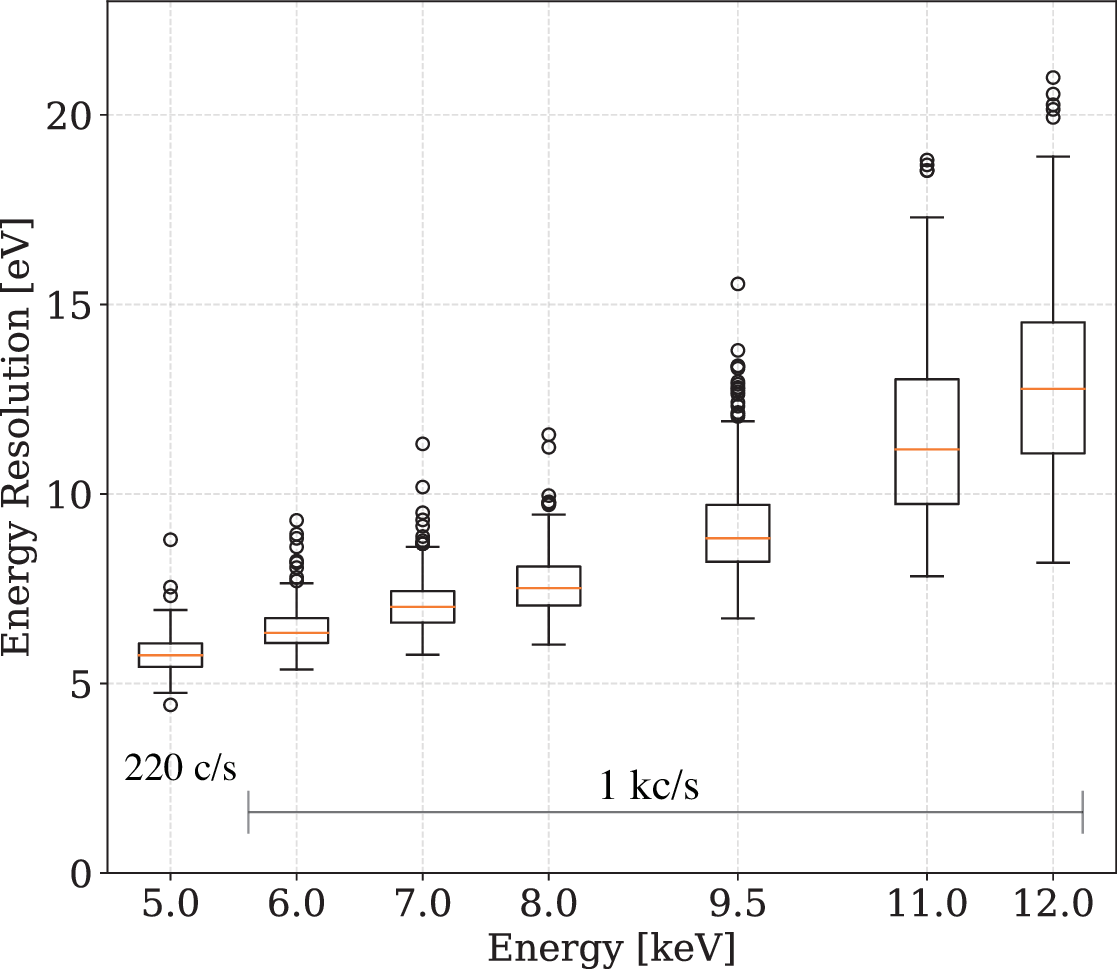}
\caption{The energy resolution at different energies obtained by elastic scattering from a Pb target. 
The box extends from the lower to upper quartile values of the energy resolutions in 220 pixels, 
with a horizontal line at the median. 
The upper whisker extends to last datum less than $Q_3$ + $R\times 1.5$, 
while the lower whisker extends to the first datum greater than Q1 - $R\times 1.5$, 
where $Q_3$ and $Q_1$ are the upper and lower interquartile, and $R = Q_3-Q_1$. 
Beyond the whiskers, data are considered outliers and are plotted as individual points.  
}
\label{resol_vs_e}
\end{center}
\end{figure}

We measured the energy resolution as a function of energy using the beam. 
We used a film of Pb to elastically scatter different beam energies and obtained the energy resolution 
by fitting the line with a gaussian profile. The result is shown in figure \ref{resol_vs_e}. 
The count rates were set at $\sim$ 1 kc/s/all pixels at energies above 6 keV; 
the maximum beam intensity available at 5 keV produced a count rate of only 220 c/s/all pixels for the current setup. 
In contrast to figure \ref{resol}, the outlier values in figure \ref{resol_vs_e} could be caused by high count rates due to Bragg reflection in the materials, though this needs further investigation.

This degradation in energy resolution at higher energies is believed to be caused by non-linearity of the transition curve of the TES and their readout system. 
One non-linear effect is that the resistance of the TES is non-linear as a function of temperature, which is more significant as it gets closer to the normal resistance. The change in resistance is read out and amplified by SQUIDs, which are operated in the feedback mode called a flux-locked loop to linearize its output, however the output voltage becomes slightly non-linear as the size of the input signal increases.
These effects cause the pulse size to grow more slowly than the photon energy, which leads to a degradation in energy resolution. 
These non-linear effects also cause the pulse shape to change with photon energy. 
Since the optimal filtering algorithm is based on the assumption that the signals are proportional to energies and the noise is stationary, 
the non-linear effects in a realistic environment could violate the assumption. 
This is why the energy resolution degrades at higher energies and has to be experimentally evaluated.  

\begin{figure}[htbp]
\begin{center}
\includegraphics[width=0.99\linewidth, keepaspectratio]{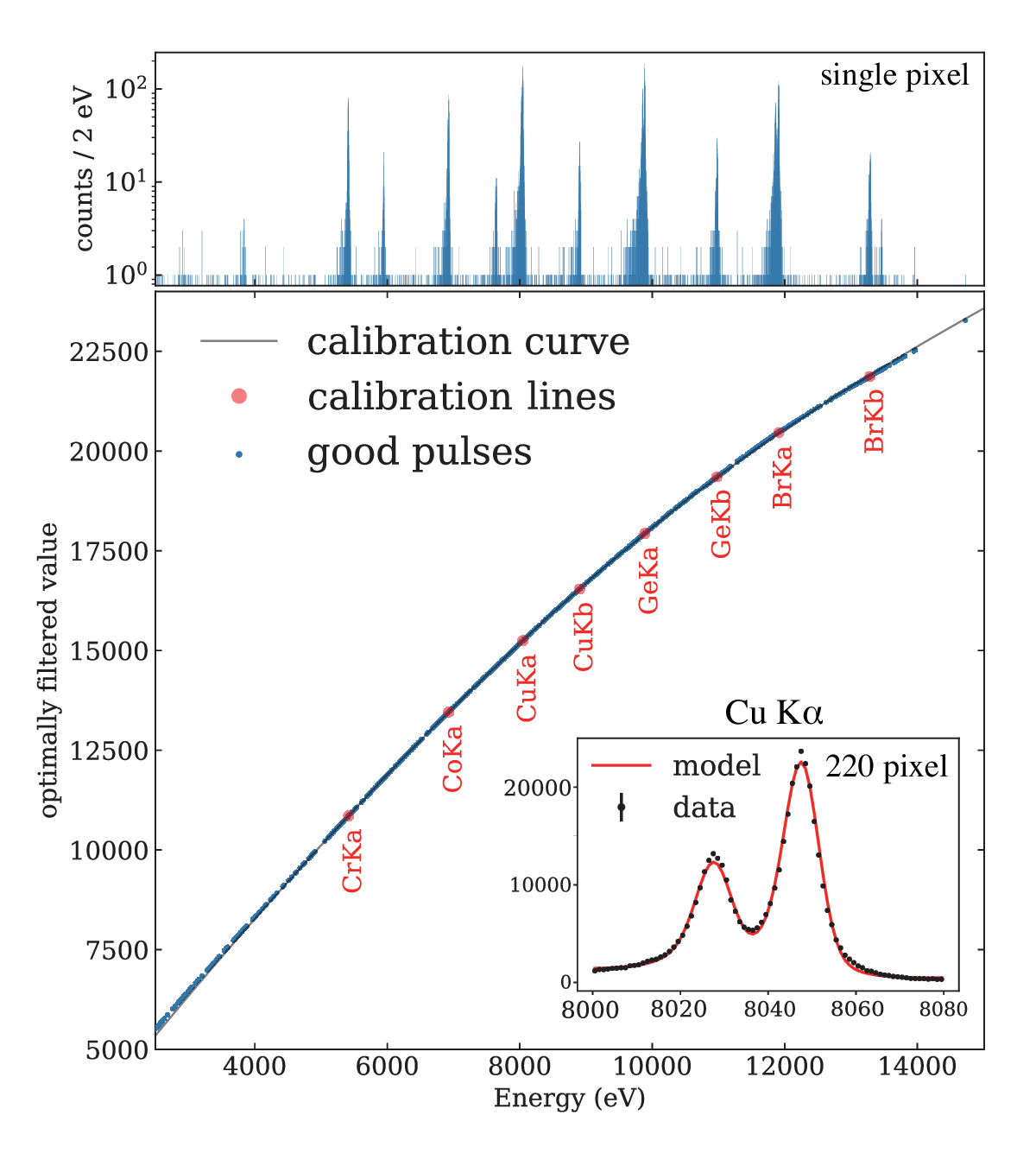}
\caption{(top) The energy spectrum of calibration sources taken from a single pixel A. (bottom) The obtained relation of pulse height and energy from pixel A. 
The red points are emission lines from Cr, Co, Cu, Ge, and Br, and the approximate calibration curve is overlaid. 
The inset shows, for the Cu K$\alpha$ line, the summed spectrum over 220 pixels (black points) and the fitted line profile (red solid). 
}
\label{ecal}
\end{center}
\end{figure}

\subsection{\label{sec:level2-1}Energy calibration}

The pulse-height and energy relation is determined 
with Cr K$\alpha$, Co K$\alpha$, Cu K$\alpha$, Cu K$\beta$, Ge K$\alpha$, Ge K$\beta$, Br K$\alpha$, and Br K$\beta$. 
They are generated by fluorescing a pellet of Cr$_2$O$_3$, CoO, CuO, GeO$_2$, and KBr using a supporting agent of boron nitride 
with a beam energy of 14 keV.
The example of an energy scale from pixel A is shown in figure \ref{ecal}. 
The data points are fitted with a third-order B-spline in log-log space, which does not pass through the origin.
The nonlinearity of the TES response becomes more pronounced at higher energies, 
due to the TES's approach to its normal-state resistance at the top of higher-energy pulses.
We created energy calibration curves for each pixel at every ADR recycle. 
The gain of the system drifts by $\sim$0.01\% over $\sim$10 hours. 
It is caused by the variation in the temperature of the second stage of the ADR from roughly 500 mK to 1 K over 24 hours 
which affects the gain via multiple independent mechanisms. 
%Since the gain of the output is controlled by series array SQUIDs installed on the 
%500mK plate, and the temperature of this plate gets higher as the current of the ADR decreases, there is additional gain drift over the course of an ADR cycle that must be corrected in pulse processing. 
The gain drift is strongly correlated to the pretrigger (256 samples before the trigger), 
and thus most of the correlation between pulse height and the pretrigger is removed by the ``drift correction'' step\cite{Fowler2016}.
%Before this correction, the dependence may be $\sim$2 eV per 100~mK increase at the 500 mK plate, which typically takes several hours but could depend on thermal environment and a setup. 
After this correction, the effect is reduced to a scale of about 1~eV over 10 hours after the ADR recycle, and can often be further reduced with a correction based on a slowly varying function of time. If we assume that this effect adds in quadrature with other sources of the energy resolution, then we might expect it to degrade a 10 hour long co-added spectrum 
from a detector with 5 eV resolution to one with 5.1 eV resolution.  
More careful treatment is required for longer measurements, or when the absolute energies of emission lines is a goal of the measurement.
In this application, a typical measurement duration is less than two hours, so the effect of the gain drift is small. 
Additionally, the dependence of filtered pulse heights on the exact pulse-arrival time needs to be corrected\cite{Fowler2016}, 
though arrival time contributes less than does gain drift to the energy resolution because the sampling rate is fast enough to track the rising profile of the pulse.  
After applying these corrections, converting pulse heights to energy via the best-fit calibration curve for each pixel, 
and adding the histogrammed counts for all pixels, 
we obtain the final emitted X-ray spectrum. 
The inset to figure \ref{ecal} shows the Cu K$\alpha$ line, 
which is well-fit to the H\"{o}lzer line shape \cite{1997PhRvA564554H} convolved with a gaussian line shape and a an exponential low-energy tail.

%By applying these corrections and using the calibration curve, a combined spectrum is obtained as shown in the inset in figure \ref{ecal} and well reproduces the Cu K$\alpha$ line profile \cite{1997PhRvA564554H} convolved with energy resolution and the low energy tail.  

\subsection{\label{sec:level2-1}High rate tolerance}
%figure of dE and dead time vs. input counts 

The decay time-constant of our TESs is about 1 ms.
In order to qualitatively assess the degree of pileup, 
\begin{figure}[htbp]
\begin{center}
\includegraphics[width=0.99\linewidth]{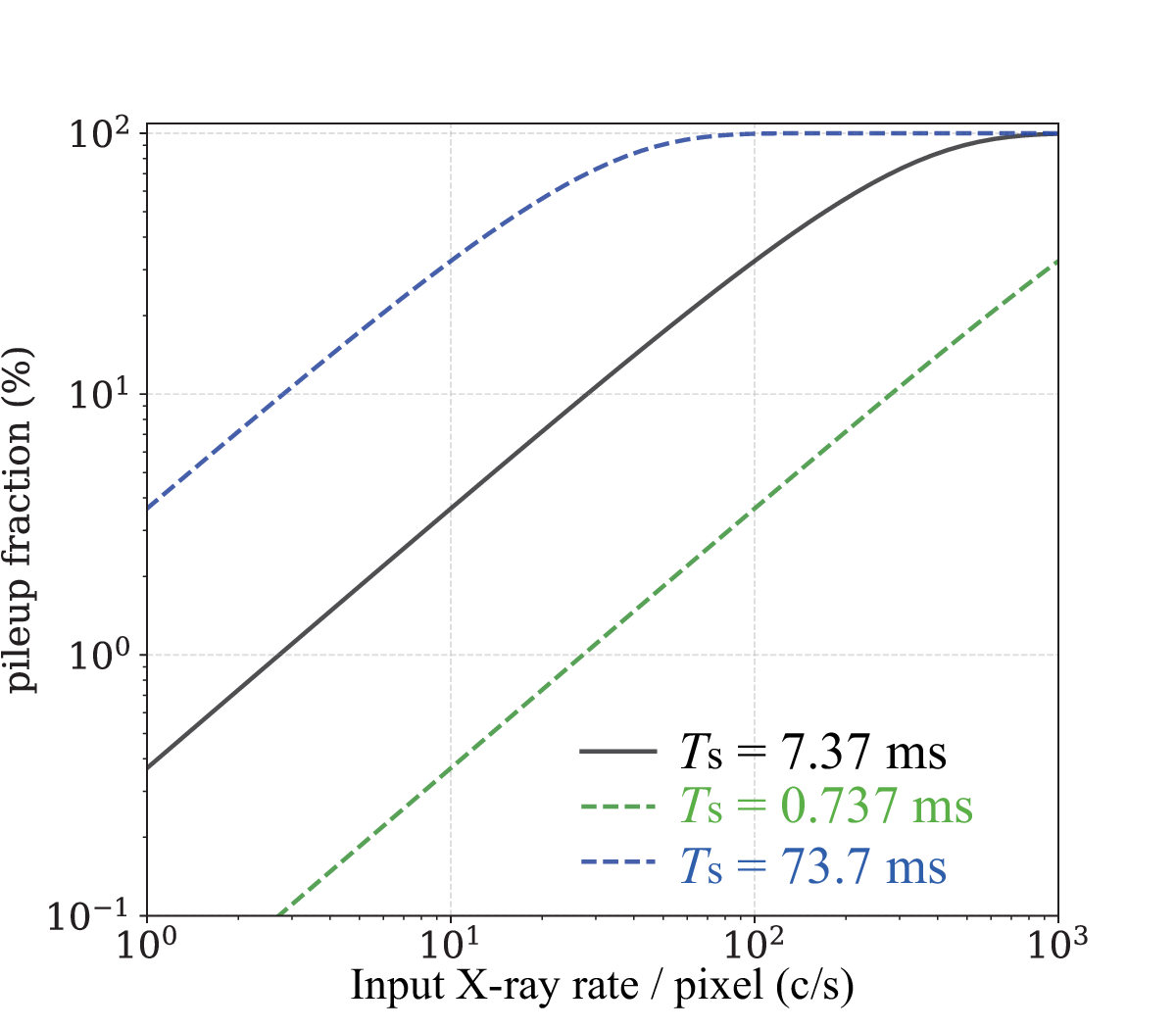}
\caption{The theoretical pileup fraction (\%) vs. incoming X-ray count rate per pixel. 
The pileup fraction for the nominal $T_{\rm{s}}$ is shown as a black solid line. 
For reference, the pileup fractions for data-record lengths that are ten times shorter and longer than the nominal length are shown in a green and blue dashed lines, respectively.}
\label{pl}
\end{center}
\end{figure}
we define the pileup fraction: $f_{\rm{pf}}(x)$ is the ratio 
of $P(k\ge2,x)$ to $P(k\ge1,x)$, 
where $P(k,x)$ is the Poisson distribution function, 
$x$ is defined as the mean number of input X-rays for a pixel per one record time $T_{s}$, 
and $k$ is the integer number of photons per $T_{\rm{s}}$. 
%Since the count rate per pixel, $r$, is often used as a variable, we relate it to $x$, as $x = r \times T_{\rm{s}}$. 
The mean number of X-rays per pixel per record time is related to the input X-ray rate per pixel, $r$, as $x = r \times T_{\rm{s}}$.
Then, the pileup fraction, $f_{\rm{pf}}(x)$, is expressed as
\begin{equation}
\begin{split}
f_{\rm{pf}}(x) = \frac{\Sigma_{k=2}^{\infty} P(k,x)}{\Sigma_{k=1}^{\infty} P(k,x)} &= \frac{1-(1+x)e^{-x}}{1-e^{-x}} \\
&=\frac{1}{2}x - \frac{1}{12}x^2 + O(x^4) \\
\end{split}
\end{equation}
The pileup fraction is plotted as a function of $r$ in figure \ref{pl} with a nominal $T_{\rm{s}}$ (=7.37~ms), as well as with an order of magnitude higher and lower values for comparison. 
When $r$ = 10 c/s/pixel, or 2200 c/s/220 pixels, 
the pileup fraction is $\sim$ 3\%; while when $r$ = 100 c/s/pixel, the pileup fraction is larger than 30\%. 
Thus, the input X-ray rate is adjusted in the range below $r \sim$ 10 c/s/pixel. 
When the pulse decay time constant is longer, such as in a gamma-ray TES, which might have a ten-times slower decay time than an X-ray TES, 
the limit would be lowered by an order of magnitude. 
In contrast, a faster TES is better unless the shorter readout time worsens the energy resolution. 

In this experiment, we discard data records in which there is more than one X-ray arrival. 
However, in some applications of microcalorimeter arrays, detection efficiency is a critical enough parameter that more effort is made to recover pile-up events.
%Although this experiment discards time periods where more than one events, when they are large enough to be detected, arrive within a frame time (=7.37~ms), 
%utilizing pile-up events is important for energy resolution and detection efficiency and there are intensive efforts in this field. 
One of the advanced examples is implemented in the onboard pulse shape processor (PSP) in the ASTRO-H SXS. 
The program in a real-time OS can subtract a template waveform from the primary pulse, and estimate the pulse shape of the subsequent event. 
Their system can detect two events that are separated by more than 2~ms, 
provided that the two adjacent pulses have a pulse amplitude contrast of a factor of 1/30 or more\cite{10.1117/1.JATIS.4.1.011217}.  
%Compared to a rise time of $\sim4$~ms at a sampling time of  80 $\mu s$ for ASTRO-H PSP, 
%the rise time of the pulse in our detector is $\sim0.4$~ms at a sampling time of 7.2 $\mu s$, 
%so the difference of the process speed is not significant and thus it is possible to implement the function of searching for secondary events in our framework; 
%cf., a method of detecting nearly-coincident events using a similar detector \cite{Alpert2016}. 
The ratio of the rise time of an X-ray event to the sampling time sets the calculation speed of this triggering algorithm. 
This ratio is similar in ASTRO-H ($\sim$4~ms to 80 $\mu s$) and our detector ($\sim0.4$~ms to 7.2 $\mu s$), 
so the ASTRO-H algorithm would work for our detector.  Another method for the detection of nearly coincident events is described\cite{Alpert2016}.
However, the energy resolution of the secondary events will deteriorate, due to both the non-linearity of the pile-up X-ray pulses and inaccuracy of the subtraction of the primary X-ray pulse. 
To put the secondary-event analysis into practical application, a significant effort such as was done for ASTRO-H SXS\cite{10.1117/1.JATIS.4.2.021406} is necessary. 
We discarded pile-up records to simplify the analysis, but further studies on the pulse processing could improve our detector's throughput.

\begin{figure}[t]
\begin{center}
\includegraphics[width=0.99\linewidth, keepaspectratio]{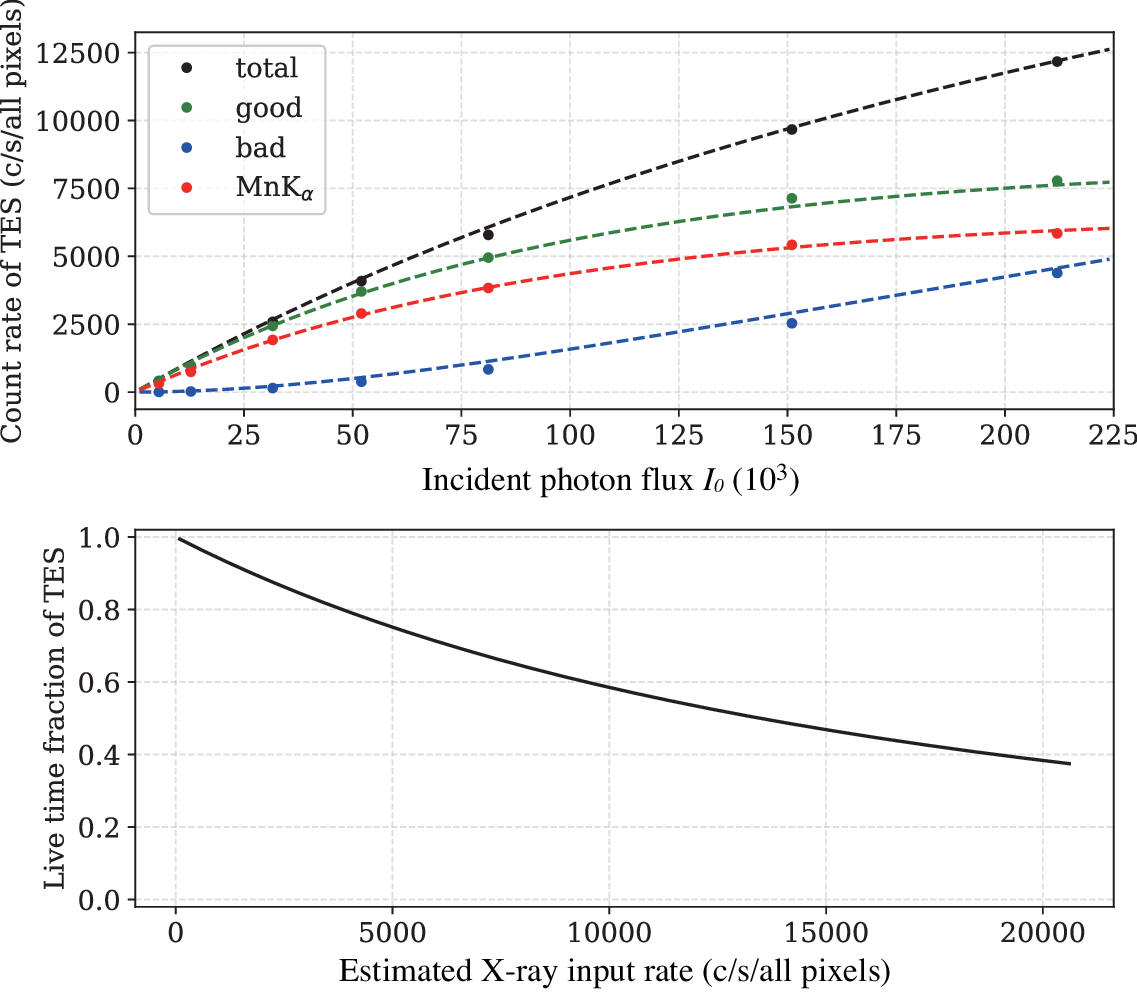}
%/Users/syamada/work/ana/from_old_mac/HEATES/SPring-8_201907/plot_rate/plot_rate_th.py 

\caption{(top) 
The array rate of data records of various types vs. the flux of X-rays on the sample measured with the ion chamber: 
the total, good, bad, and Mn K${\alpha}$ record rates are shown in black, green, blue, and red, respectively. 
The $r'_{\rm{total}}$,  $r'_{\rm{good}}$, $r'_{\rm{bad}}$, and  $r'_{\rm{Mn}}$ are overlaid in a dashed line with the corresponding color.  
(bottom) The live time fraction, $f_{\rm{live time}}$, is plotted as a function of an estimated X-ray input rate per TES, $r$. }
\label{de_det}
\end{center}
\end{figure}

In this experiment, we evaluated the performance of the detector vs. the input X-ray count rate. 
A film of MnO$_2$ was illuminated by a 6.6 keV beam, with seven different beam intensities that produced rates of triggered data records, 
summed over the 220-pixel TES array, of between 400 c/s and 12,000 c/s.
Figure \ref{de_det} (top) plots the rate of triggered data records vs. the flux of X-rays incident on the sample.
We sorted the data records into two categories: good and bad. 
Data records in the ``good" category contain only a single X-ray arrival, 
while the ``bad'' data records include records that contain multiple piled-up X-ray arrivals, 
failed operation of the flux-locked SQUID for readout of TES, and any other anomalously-shaped pulses. 
%The number of total pulses is the sum of both categories. 
The ``total'' record rate is the sum of the good and bad record rates. 
As the input count rate increases, the fraction of bad data records increases. 
The plot also shows the rate of good data records in an energy window around the Mn K${\alpha}$ line.
%At the lowest incident photon flux, about 80\% X-ray events come from Mn K${\alpha}$.% 
%though its fraction decreases almost linearly as the input flux increases. 

%We assume $r = p  I_0$ as an estimated count rate of the TES assuming that it can process all events, 
%where $p$ is a coefficient of the proportionality. 
The true input count rate of X-rays per TES pixel is $r = p  I_0$, where $I_0$ measures the beam flux incident on the sample via the current in an ion chamber and $p$ is a proportionality coefficient.
Our spectrometer does not measure $r$ directly. Instead, we measure the rates of triggered data records, $r'$. 
The record rates of good, bad, and Mn K${\alpha}$ are defined as $r'_{\rm{good}}$, $r'_{\rm{bad}}$, and $r'_{\rm{Mn}}$. 
We phenomenologically estimated the detection efficiency by modeling the obtained data in figure \ref{de_det} (top)
using the following equations: 
\begin{eqnarray}
r'_{\rm{total}} &=&  \frac{1}{1+T_s r} r \equiv Dr  ~~~~~  \rm{(c/s/pixel)}   \\
r'_{\rm{good}} &=&  D^2 r                                    ~~~~~~~~~~~~~~~~~~~  \rm{(c/s/pixel)}  \label{eq:good} \\
r'_{\rm{bad}} &=&  r'_{\rm{total}} - r'_{\rm{good}} ~~~~~~~~  \rm{(c/s/pixel)}   \\
r'_{\rm{Mn}} &=& k r'_{\rm{good}}                         ~~~~~~~~~~~~~~~~~  \rm{(c/s/pixel)}   
\end{eqnarray}
where $r = 9.2 \times 10^{-2} I_0$ and $k = 0.78$, whose coefficients are obtained by comparing the data in figure \ref{de_det} with the equations. 
They are overlaid in figure \ref{de_det} (top). 
%Figure \ref{de_det} (bottom) shows the fraction of good events to the estimated total count, so-called live time fraction, $f_{\rm{live time}} = r'_{\rm{good}}/r$. 
Figure \ref{de_det} (bottom) shows the live-time fraction, $f_{\rm{live time}} = r'_{\rm{good}}/r$, or the fraction of input X-rays that produce good data records vs. the estimated X-ray input rate over the TES array. The dead time fraction, a fraction of time during which 
TES pixel cannot write a data record that contains only one X-ray arrival, is written as 1 - $f_{\rm{live time}}$.

% (1) input rate vs. deadtime

\begin{figure}[b]
\begin{center}
\includegraphics[width=0.99\linewidth, keepaspectratio]{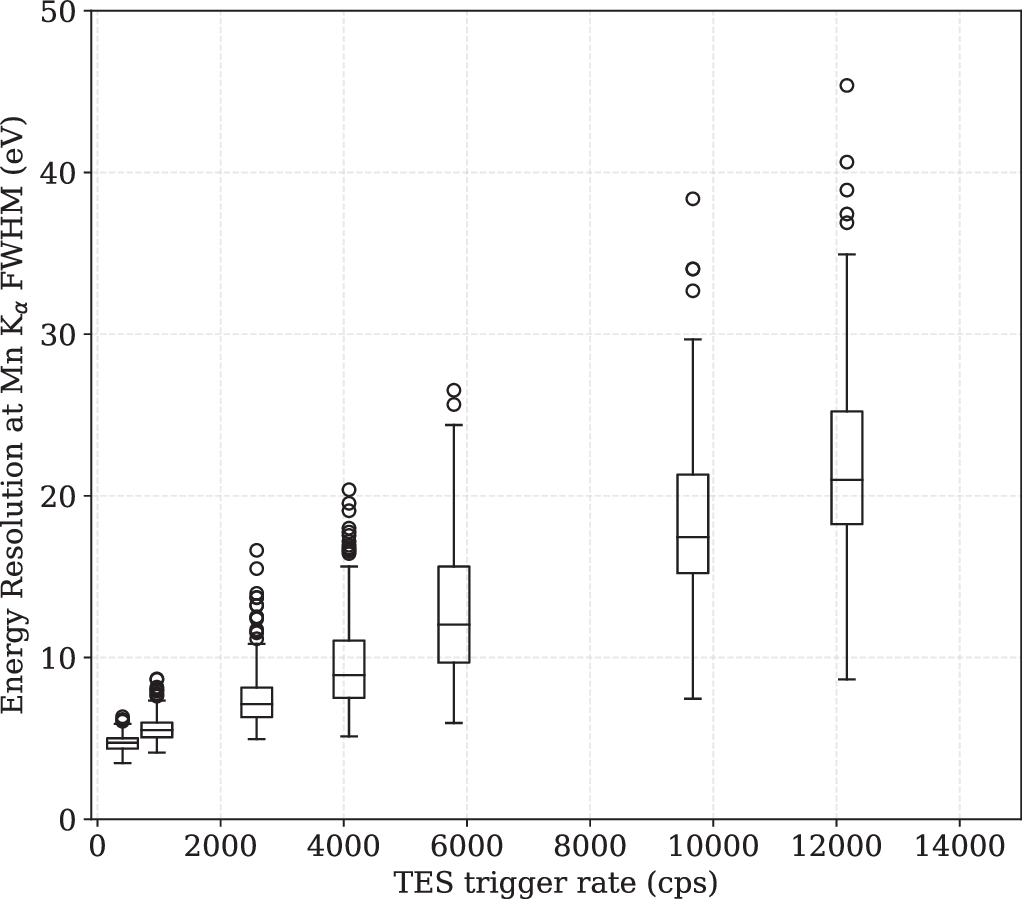}
\caption{
Energy resolution at the energy of Mn K$\alpha$ line 
against the TES trigger rates of all pixels (= the total count rate of all TES pixels). 
The notation of the plot is same as in figure \ref{resol_vs_e}. }
\label{in_de}
\end{center}
\end{figure}

High count rates also affect the energy resolution. 
Figure \ref{in_de} shows the energy resolution as a function of the TES trigger rate. 
The energy resolution is obtained by removing the bad events and then fitting the spectrum of each pixel to the line shape of Mn K${\alpha}$\cite{1997PhRvA564554H}. 
The energy resolution worsens with increasing count rate due to crosstalk among the TES pixels,
%A slight increase of the count rate does worsen the mean of the energy resolution and widen the spread of its distribution. 
%This is because there is known to be crosstalk among the pixels, 
which was not removed in this analysis.
A shorter record-length could improve the energy resolution 
at higher count rates, 
though an energy calibration would need to be carefully performed\cite{Tatsuno2020}. 

% (2) input rate vs. energy resolution 

\section{\label{sec:level3}Results}

\subsection{\label{sec:level3-1} Simultaneous multi-element analysis}

\begin{figure}[htbp]

\begin{center}
\includegraphics[width=0.99\linewidth, keepaspectratio]{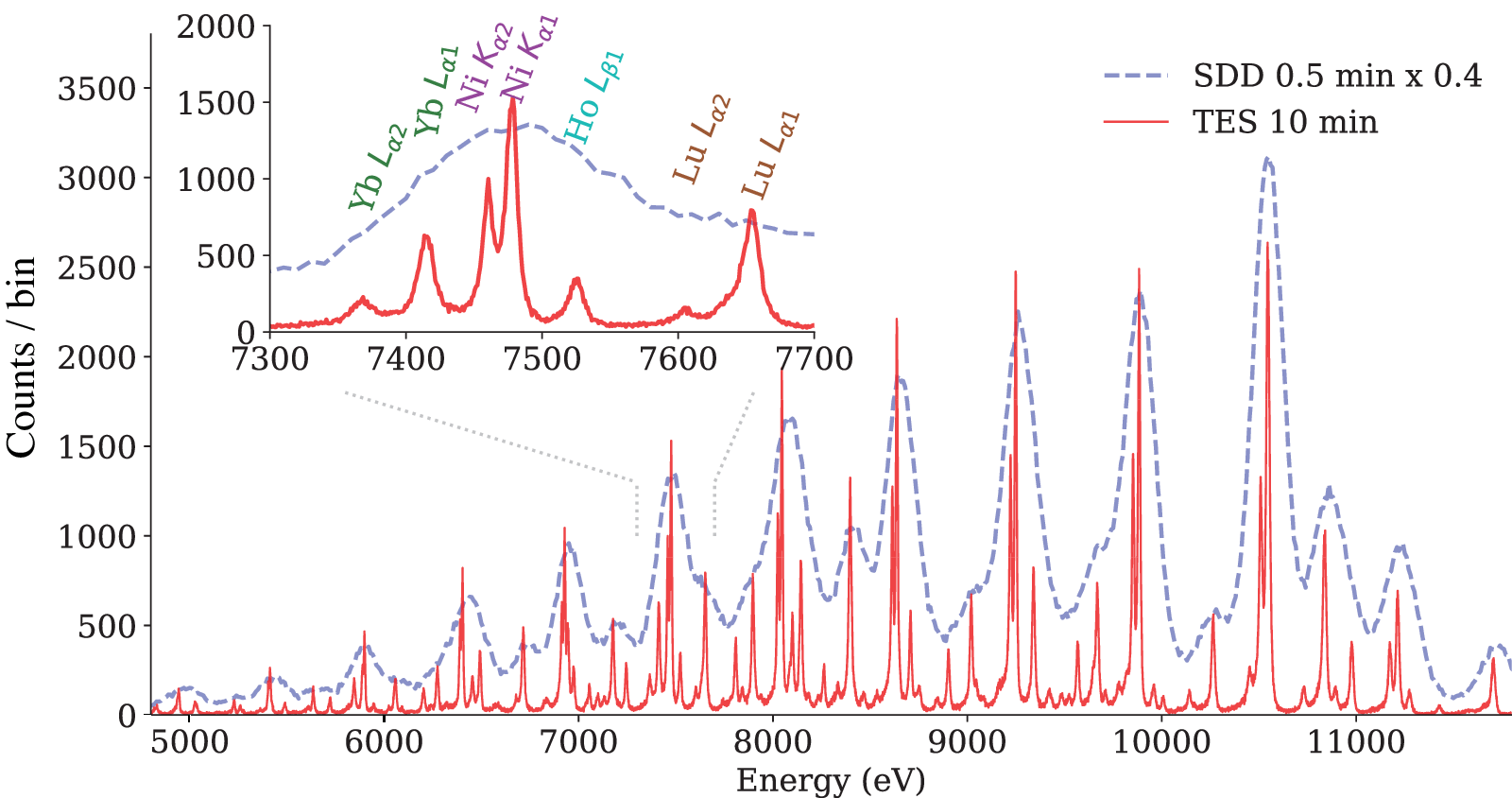}
\caption{The wide-band spectra of SRM610 taken with TES and SDD. 
The binsize for the TES data and the SDD data are 1~eV and 10~eV, respectively. 
The energy of the X-rays illuminating the target was 14 keV. 
The TES spectrum was created by summing over all pixels and integrating for ten minutes (solid red). 
The SDD spectrum was created from half a minute's data and overlaid in the plot by multiplying by a factor of 0.4 for visual clarity (dashed blue). The inset shows an example of the detailed comparison, where the TES can probe Yb, Ho, and Lu, which the SDD could not resolve due to the presence of strong Ni K lines.}
\label{nist610wideband}
\end{center}
\end{figure}

To demonstrate the capabilities of the TES at BL37XU in SPring-8, 
a NIST standard reference glass (NIST SRM 610)\cite{NIST610} is used as a target. Figure \ref{nist610wideband} shows the spectra of SRM610 as measured by the TES and the SDD. 
The TES spectrum is a sum of 220 pixels accumulated for 10 minutes, 
while that of the SDD is from 0.5 minutes. 
The difference in the count rates is due to geometrical effects and the materials between the detector and the target. 
The spectrum of the SDD in figure \ref{nist610wideband} is composed of several sharp peaks. 
Since the energy of K lines are approximately proportional to the square of the atomic number, $Z$, 
%the difference of the K$\alpha$ energies among the different elements in 4--13 keV is an order of hundreds of eV. 
the K$\alpha$ complexes in the energy range of 4 keV to 13 keV occur every few hundred eV.
The intensity ratios of the K$\alpha$ and K$\beta$ lines are about 10 to 1, 
though their rates are weakly dependent on the chemical composition. 
The L lines from high $Z$ elements complicate line identification in the spectrum. 
As shown in the spectra in the inset of figure \ref{nist610wideband}, 
the spectrum of the SDD around the Ni K lines appears to be a single peak, while the TES can resolve 
L lines from Yb, Ho, and Lu, in addition to the Ni K lines. 

To fit the spectrum with an estimated one (so-called model), 
the relevant physical processes including fluorescence emission, the photo-absorption, and the detector response 
are calculated to produce an expected spectrum.  
The free parameters to generate the expected spectrum are 
the gaussian energy resolution of the TES, the exponential low-energy-tail fraction of the TES response, 
the constant level of the background, and the abundance of each element in the composition of the reference glass.
The number of the free parameters is three plus the number of the elements. 
The line energies and the intensity ratios are fixed to the reference value obtained from $xraylib$\cite{SCHOONJANS2011776}. 
This simplifies a parameterization of the model, reduces the computational time, and enables us to estimate the relative abundance. 

The geometry is approximated by assuming that the sample is at 45$^\circ$ to both the beam and the TES. 
The X-ray intensity of fluorescence lines just outside the sample is calculated as follows: 
1) The incident X-ray, $E_{\rm{in}}=14$~keV, is photo-absorbed along at the depth of $x$ from the surface of the sample.
2) An X-ray fluorescence photon is emitted at $x$. 
3) The fluorescence X-rays are subject to photo-absorption within the sample itself.
4) The remaining photons can escape from the sample to the air. 
We define $\sigma^{\rm{att}}(E)$ as the photo-absorption cross sections for X-rays with an energy of $E$, 
$\sigma_{\mathrm{i}}^{\alpha}(E) $ is the X-ray fluorescence cross section for a given atomic number $i$ and 
and an excited state of $\alpha$, and $ \rho \equiv \sum_i \rho_i$ is the sum of the number density over the element $i$. 
The intensity of the fluorescence line from an element $i$ and an excited state $\alpha$ 
is a function of the energy of the input beam $E_{\rm{in}}$ and the energy of fluorescent X-ray $E_{\alpha}$, 
which is expressed as;

%\begin{equation}\label{eq:selfabs}
%\int^{d}_{0} (N_0 e^{-\sigma_{14}\rho x})
%(\sigma^\mathrm{elem,trans}_{14,\mathrm{fluo}}\rho^\mathrm{elem})
% (e^{-\sigma_{E_\mathrm{fluo}}\rho x}) dx,
%\end{equation}
%where $N_0$ is the number of incident X-ray photons, $\sigma_{14}$ and $\sigma_{E_\mathrm{fluo}}$ are total attenuation cross sections for the X-rays whose energies are 14~keV and $E_{\mathrm{fluo}}$, respectively, $\sigma^\mathrm{elem,trans}_{14,\mathrm{fluo}}$ is the X-ray fluorescence cross section for a given combination of the element and transition at 14 keV, $\rho$ and $\rho^\mathrm{elem}$ are the mean density of the sample and the density of the corresponding element, respectively. 

\begin{equation} \label{eq:selfabs}
I_i^{\alpha}(E_{\rm{in}}, E_{\alpha}) \sim \int^{d}_{0} \{N_0 e^{-\sigma^{\rm{att}}(E_{\rm{in}}) \rho x}\}
\{ \sigma_\mathrm{i}^{\alpha}(E_{\rm{in}})   \rho_i \}
\{ e^{-\sigma^{\rm{att}} (E_\alpha) \rho x}\} dx,
\end{equation}
where $N_0$ is the number of incident X-ray photons, and $d$ is the depth of the sample. 
$\Omega/4\pi$, the solid angle of the detector from the sample, is included if the absolute value of the count rate is needed. 
The first factor of the integrand corresponds to the number of 14~keV X-ray photons which reach a depth of $x$ from the sample surface. 
The second factor represents the amount of X-ray fluorescence induced by the 14~keV X-ray within the depth of $x\sim x+dx$. The escape probability of fluorescence X-ray photons is calculated using the third factor.

%NIST610 just for presentation

By performing the integral, Equation~\ref{eq:selfabs} reduces to
\begin{equation} \label{eq:selfabs2}
N_0 \dfrac{\sigma_\mathrm{i}^{\alpha}(E_{\rm{in}}) \rho_i}{(\sigma^{\rm{att}}(E_{\rm{in}})+\sigma^{\rm{att}} (E_\alpha))\rho}[1-e^{-(\sigma^{\rm{att}}(E_{\rm{in}})+ \sigma^{\rm{att}} (E_\alpha))\rho d}].
\end{equation}
This represents the relative intensity of each transition. 
Since the thickness of SRM610 is about 1 mm and its density is 2.65 g/cm$^3$ according to \cite{B917261K}, 
% https://pubs.rsc.org/en/content/articlelanding/2010/ja/b917261k#!divAbstract
% [syamada] $ python                                              [~/work/ana/sp8tes/nist610/calc_abundance_for_paper/check_cs]
% import xraylib
% import math
% rho = 1 # g/cm^3
% d = 0.1 # mm
% math.exp(-xraylib.CS_Total_Kissel(26,6.0)*rho*d) 
% 0.00021279157039573232
% math.exp(-xraylib.CS_Total_Kissel(26,14)*rho*d)
% 0.0010137910018029883
the exponential term can be ignored because it becomes almost zero; 
e.g., when the total (i.e., photoionization, Rayleigh and Compton scattering) attenuation cross section of iron $\sigma_{Fe}(E)$  $\rm{[cm^2/g]}$ as a function of an energy $E$ is considered for explanation, 
the exponential terms become $e^{-\sigma_{Fe}(6\rm{keV})\times0.1 \rm{[g/cm^2]}} \sim 0.0002$ and $e^{-\sigma_{Fe}(14\rm{keV})\times0.1 \rm{[g/cm^2]} } \sim 0.001$. 
The intensities of all the transitions are calculated and summed up, to obtain the resulting model spectrum. 
Here the model spectrum means that the spectrum of X-rays escapes from the SRM610 target.
The abundance of the elements, not the amplitudes of the individual lines, is floated in the fit. 
%The energy resolution and the tail fraction of the TES, and the normalization factor of the constant background are floated in the fits. 
The transmission of the Al and Be filters, air, and the photo-absorption efficiency of the Bi absorber are 
taken into account as shown in figure \ref{tesqe}. 
Note that the scattered photons in the air is not negligible in the background spectra. 
\begin{figure*}[htbp]
\begin{center}
\includegraphics[width=0.92\linewidth, keepaspectratio]{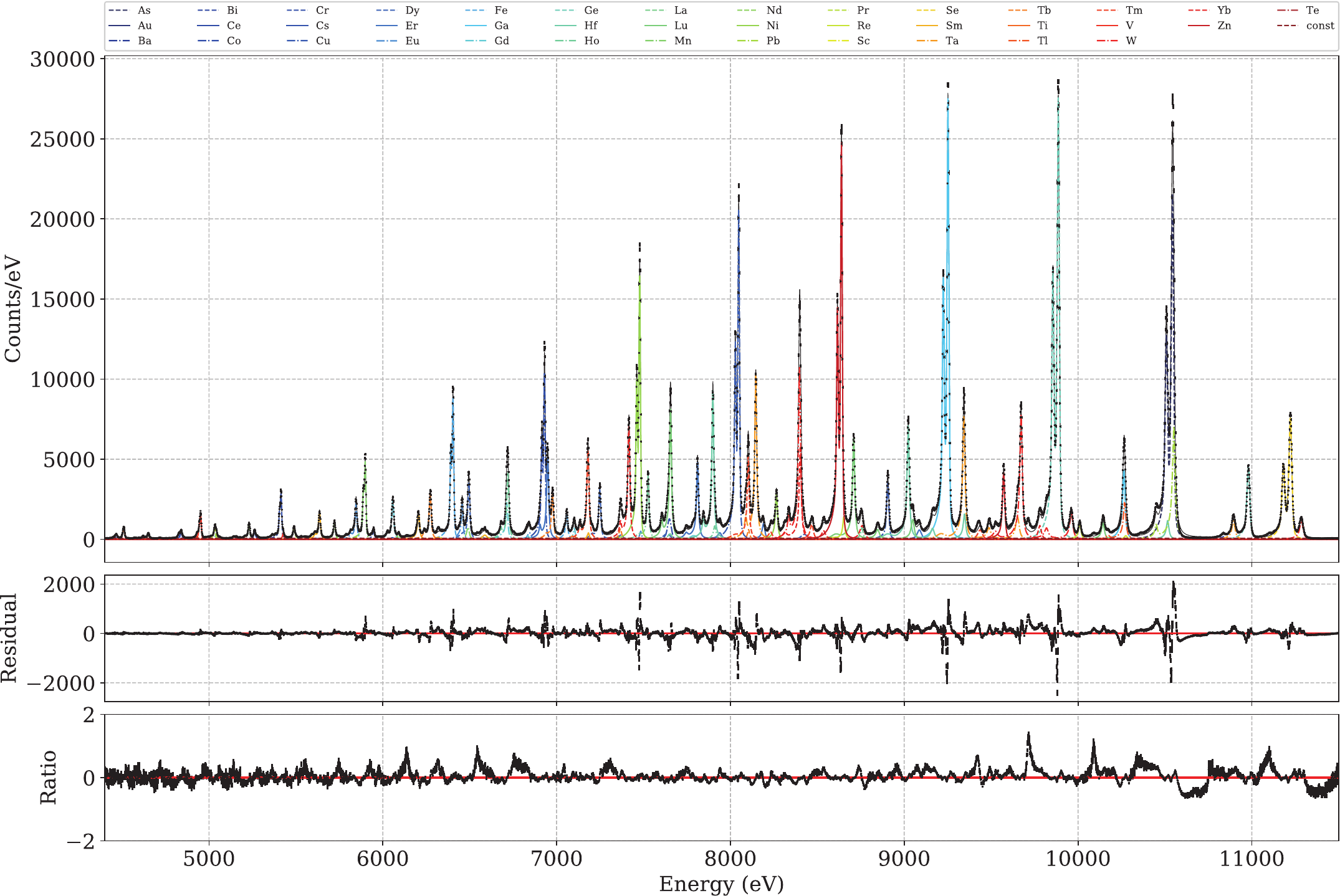}
%/Users/syamada/work/ana/sp8tes/nist610/calc_abundance_for_paper/check_Ba_narrow/fig_paper_final/sp8_fitting_tailfit_natural_glass_paper_finalfig.py
\caption{(top) The example of simultaneous multicomponent fits for the XES of TES. The XES of TES and the best-fit model components are shown. 
Each element is plotted in different colors and line styles. (middle) The residual between the data and the model. (bottom) The ratio of the residuals, which is (data - model)/model. } 
\label{nist610fit}
\end{center}
\end{figure*}

We fit the spectrum with the sum of the lines of known elements. 
The fit is done by comparing the model to the obtained spectrum and 
looking for the minimum of the residuals by changing the model parameters. 
The best-fit model and the residuals between the data and the model are shown in figure \ref{nist610fit}. 
There still remains some residuals between the spectrum and the model, which are probably due to uncertainties in the detector response 
and the line shape of some of the emission lines. 
Since the relative statistical errors in the spectrum are smallest near the peaks and are larger in the tails, 
the fitting, to minimize the residuals divided by the statistical errors, tends to converge to obtain the line intensities. 
%The line intensities obtained by the fits are well determined since the peaks of the lines are most dominant in the fitting process of minimizing 

\begin{figure}[htbp]
\begin{center}
% cp ~/work/ana/sp8tes/nist610/calc_abundance_for_paper/check_Ba_narrow/evaluate_all_errors/nist610_for_paper_fit_global_local.eps .
\includegraphics[width=0.95\linewidth, keepaspectratio]{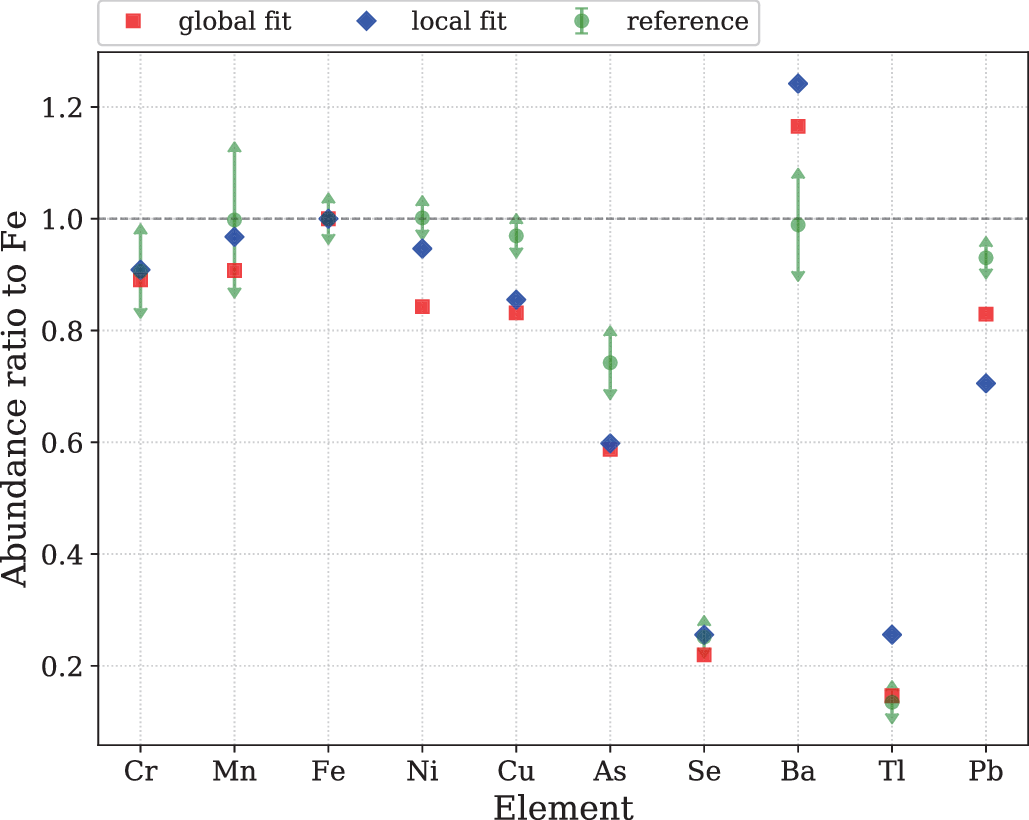}
\caption{The abundance ratios to Fe obtained by fitting the TES spectrum including the quantum efficiencies in figure \ref{tesqe} and the self absorption in eq.~(3) 
are shown in red (global fit) and blue (local fit), respectively. 
The reference values are shown in green, where the uncertainties in the reference are plotted with arrows.
} 
\label{fitpara}
\end{center}
\end{figure}

The resultant abundance ratio as a reference to Fe obtained from the best-fit model in figure \ref{nist610fit} is shown in figure \ref{fitpara}.
The large number of photons creates small statistical errors, 
so the uncertainties of the fit is dominated by a systematic error. 
To obtain a rough estimation of systematic uncertainties, 
several representative lines are individually fitted in a narrow energy range around each line center. 
The results of the local fits are plotted in figure \ref{fitpara}. 
The difference of the results between the global fit and the local fits gives a rough reference on the level of uncertainties. 
The reference value and its uncertainty of SRM610 are also plotted in figure \ref{fitpara}. 
The fitting results are more or less consistent with those in the reference. 
The merit of the TES is to measure both light and heavy elements simultaneously in the hard X-ray band. 
Furthermore, resolving the lines can reduce the risk of misinterpreting the lines, 
and hence the accuracy of the measurement is improved. 

\subsection{\label{sec:level3-2} XANES of heavy elements -- As and Pb}

Mapping a local area at the $\mu$m scale using XRF with a focused micro-X-ray beam is an important method to determine distribution of trace and toxic elements in the ground, plants, and other forms of life, since the method can be directly applied to these samples under ambient conditions. 
In particular, coupling XRF with X-ray absorption fine structure (XAFS) spectroscopy has made it a powerful tool, since we can obtain species of the elements, essential information to understand the environmental behavior of toxic elements. Among the various toxic elements, arsenic (As) and lead (Pb) are the most important elements in terms of their health risks and wide occurrence in the earth's crust (e.g., \cite{Florea2006}). These two elements frequently co-exist in specific environments such as in sulfide mine tailings and coal  (e.g., \cite{Davenport1991}), which warrants studying the environmental chemistry of the two elements in the same systems (Brown et al., 2011\cite{doi:10.2134/jeq2011.0297}; Liu and Luo, 2019\cite{Liu2019}). Lead has been also important in terms of its atmospheric pollution as a potential risk to health\cite{acp-7-3195-2007}. However, determination and speciation of the two elements in the same sample can sometimes be challenging when using XRF and XAFS due to the difficulty in the separation of As K$\alpha$ and Pb L$\alpha$ emissions, since the energy resolution of an SDD is not sufficient to resolve the lines.

\begin{figure}[htbp]
\begin{center}
\includegraphics[width=0.98\linewidth, keepaspectratio]{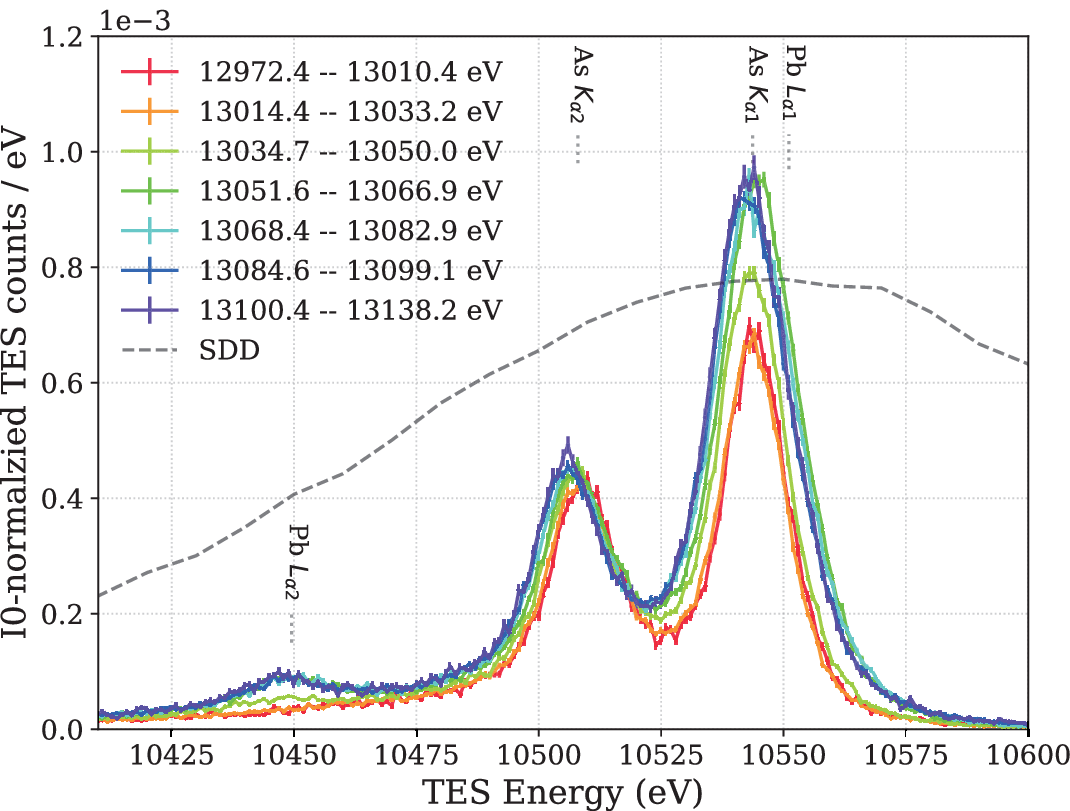}
\caption{X-ray Emission spectra of the TES spectrometer with seven different X-ray energies illuminating a sample of SRM610 across the Pb L$_{3}$-edge (13035~eV). 
The XES are the seven snap shots of continuous data taken with TES when the bean energy sweeps from 12970.3 eV to 13138.2. 
The energy scale of TES slightly varies from scan to scan due to calibration uncertainties and remaining gain drift even after a routine drift correction. 
When the input X-ray energy is below the Pb L$_{3}$-edge, only As K${\alpha1}$ and K${\alpha2}$ are present. 
As the energy increases, Pb L${\alpha1}$ and L${\alpha2}$ become prominent. As a reference, the spectrum as measured by an SDD 
when the input X-ray energy is above the Pb L$_{3}$-edge is overlaid with a dashed line.} 
\label{aspbspec}
\end{center}
\end{figure}

To overcome the difficulty of using an BCLA, XRF and XAFS using a TES is an important option. We scanned SRM610 across the Pb L$_{3}$ edge to demonstrate how the TES can contribute to the chemical diagnostics of heavy elements in complicated compounds. Figure \ref{aspbspec} shows the seven snap shots of the TES spectrum of As and Pb when the beam energy sweeps from 12970.3 eV to 13138.2. The sweep time is 60 sec over the Pb L$_{3}$-edge energy, with 30 sec in the pre/post-edge region. There are four emission lines in this band; 10508.0, 10543.7, 10449.5, and 10551.1 eV, corresponding to As K$\alpha$2 and K$\alpha$1, Pb L$\alpha$2, and L$\alpha$1.

\begin{figure}[htbp]
\begin{center}
%python  sp8_comp_PbAs_10400-10475_paper_with_refxanes.py       ~/work/ana/from_old_mac/HEATES/SPring-8_201907/PbA
\includegraphics[width=0.98\linewidth, keepaspectratio]{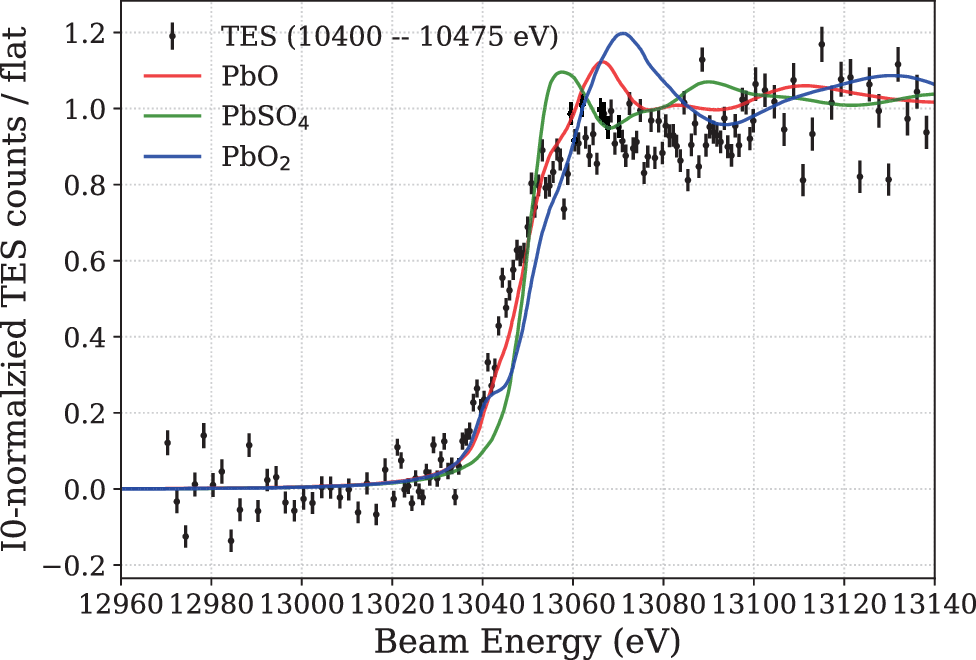} % RUN143 is used, not RUN142
\caption{The XANES of Pb L${\alpha2}$. The counts in 10400--10475~eV of TES are divided by $I_0$ counts, 
the mean below the pre-edge is subtracted, and the result is normalized at the post-edge (flat) region.
The error bars refer to the statistical errors. The XANES taken from reference samples of PbO, PbSO$_4$, and PbO$_2$, are overlaid in red, blue, and green, respectively.}  
\label{Pbla2}
\end{center}
\end{figure}

When using an SDD, the Pb L$\alpha$ and As K$\alpha$ lines may severely interfere each other, which sometimes prevents Pb mapping using L$\alpha$ in the presence of abundant As (e.g., Liu and Luo, 2019\cite{Liu2019}). In this case, Pb L$\beta$ can be used for the speciation of Pb by sweeping X-rays around the Pb L$_{2}$-edge. However, the Pb L$_{2}$-edge energy (15200 eV) is exactly identical to that of Rb K-edge (15200 eV), both of which are commonly found in relatively high concentration (average concentrations in the upper continental crust: Rb and Pb are 112 mg/kg and 20 mg/kg, respectively; Taylor and McLennan, 1985\cite{Taylor1985}), which can cause another interference. Hence, it is prudent to focus on Pb L$\alpha$2, which is $\sim$ 50 eV away from As K lines, to map Pb by XRF. Figure \ref{Pbla2} shows XANES of the Pb L$_{3}$-edge (13035~eV) measured by the TES. The statistical errors are calculated from the Poisson error of the number of counts in one step in incident beam energy. The counts are divided by $I_0$ during each step and subtracted from the mean before the pre-edge region, and the XAFS spectrum is normalized at the post-edge region, the so-called ``flat'' region. The same manner is applied to other XAFS spectra. 
The detection limit of Pb L$\alpha$2 depends on the statistical uncertainty rather than systematic uncertainties on the contamination of neighboring lines.
The selective detection of Pb L$\alpha$2 is difficult using an SDD due to its low energy resolution. As shown in figure \ref{aspbspec}, 
the XRF spectrum measured by SDD is a smeared sum of four peaks of As K$\alpha$1, As K$\alpha$2, Pb L$\alpha$2, and Pb L$\alpha$1. 

Pb L$_{3}$-edge XANES is sensitive to the valence state of Pb, either Pb(II) or Pb(IV), and its coordination environment.
The XANES of PbO, PbSO$_4$, PbO$_2$ are overlaid as references in figure \ref{Pbla2}. 
The presence of any peaks or shoulders in the pre-edge region 
(small shelf in the blue line at 13040 eV in figure \ref{Pbla2}) 
is indicative of Pb(IV) due to the electron transition from 2p to 6s, since the 6s orbital is empty for Pb(IV). 
The absence of such characteristics in the spectrum in figure \ref{Pbla2} suggests that the 6s orbital is filled in this sample. Thus, the Pb in the sample is likely to have its origin in Pb(II).

The use of TES enables us to measure Pb L$_{3}$-edge XAFS by detecting Pb L$\alpha$2 
when the statistical errors from the tails of As K$\alpha$ lines and the background is smaller than the signal from Pb L$\alpha$2. 
Note that the speciation of As using its K-edge XAFS does not interfere with Pb, since the absorption edges of Pb are higher than As K-edge.  
However, there is huge demand for the speciation of Pb in the presence of As. 
Thus fluorescence XAFS using a TES is worth considering as an option for a next-generation synchrotron facility.

%There are two known harmful elements; As and Pb. 
%Their localization in the ground, the plant, and the creature can affect our environment. 
%In recent years, atmospheric Pb concentrations have decreased due to the use of lead-free gasoline. 
%However, atmospheric pollution by Pb is still being recognized as a potential risk of the health\cite{acp-7-3195-2007}. %(Murphy et al., 2007). 
%The separation of As K$\alpha$ and Pb L$\beta$ has been difficult since the energy resolution of SDD is not enough to resolve the lines. 
%The characterization of Pb species in size-fractionated aerosols under little contamination of As was first determined for the first time with XAFS spectroscopy at SPring- 8 and the KEK Photon Factory Advanced Ring (PF-AR)\cite{KoheiSakata201720456}. 
%Yet, the separation of Pb from As is regarded as one of the remaining issues in environmental chemistry. 
%
%We scanned NIST610 across the Pb L$_{3}$ edge to demonstrate how the TES can contribute to the chemical diagnostic of the heavy elements in the complicated compounds. 
%Figure \ref{aspbspec} shows the TES spectrum of As and Pb when the beam energy sweeps from 12950.0 eV to 13114.95. 
%The sweep time is 60 sec over the edge energy, during 30 sec in the pre/post-edge region. 
%There are four lines in this band; 10508.0, 10543.7, 10449.5, and 10551.1 eV 
%corresponds to As K${\alpha2}$ and K${\alpha1}$, Pb L${\alpha2}$, and L${\alpha1}$. 

%figure of setup
%python sp8_comp_PbAs_forpaper.py         [~/work/ana/from_old_mac/HEATES/SPring-8_201907/PbAs

%Pb L${\alpha2}$ is free from the contamination from As K lines. 
%The XAFS spectrum is created by using the counts of Pb L${\alpha2}$ and shown in figure 
%\ref{Pbla2}. The statistical errors are calculated from the Poisson error of the counts in one step. 
%The counts are divided by $I_0$ during each step, and the XAFS spectrum is normalized at the post-edge region.  Therefore, its detection limit of the lines depends on the statistical uncertainty rather than systematic uncertainties on the contamination of neighboring lines. 

%    #pbla2 : 10449.5keV, Pb, LA2 11
%    #pbla1 : 10551.1keV, Pb, LA1 100
%    #aska2 : 10508.0keV, As, KA2 51
%    #aska1 : 10543.7keV, As, KA1 100  

\subsection{\label{sec:level3-3} XANES of a dilute sample -- Fe}

\begin{figure*}[htbp]
\begin{center}
%/Users/syamada/work/ana/from_old_mac/HEATES/SPring-8_201907/Fe_Kurisu/sp8_comp_Fe_paper_finalwithSDD.py
\includegraphics[width=0.95\linewidth, keepaspectratio]{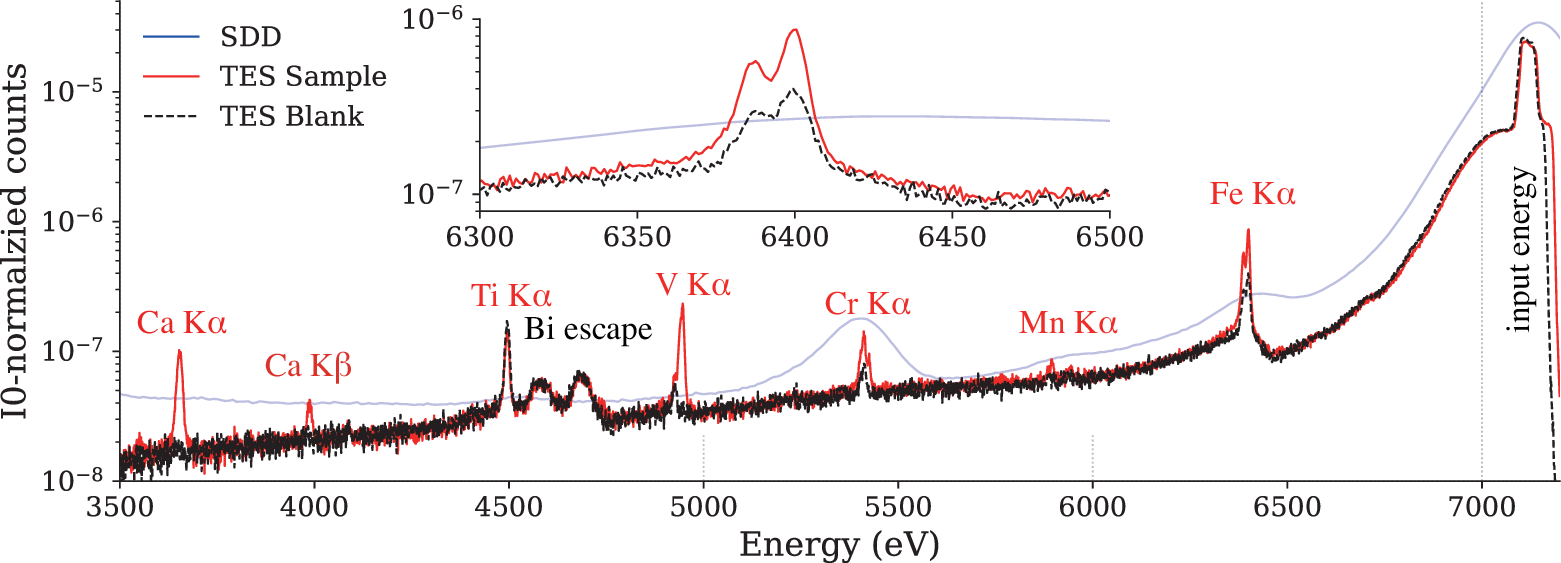}
\caption{The XES of TES taken from the aerosol sample (red) and the blank target (black). The data are accumulated during the period of the input X-ray sweeping from 7086.9--7180.4 eV for the aerosol and 7087.2--7165.2 eV for the blank target, respectively.   For comparison, the XES of SDD taken from the same sample is plotted in blue  
when the beam energy is fixed at 7131.58 eV.  
The vertical axis of TES is the counts per 1eV bin divided by the $I_0$ intensity, 
while that of SDD is multiplied by an arbitrary factor for clarity. The inset shows a magnified view at the Fe K${\alpha}$ lines.
 } 
\label{fespec}
\end{center}
\end{figure*}
Another application of the TES is to measure samples where the material of interest is very dilute. 
The high resolution can help to resolve the signature of the elements of interest from the background and mitigate risk of misinterpreting the spectra. 
Here we chose a sample containing Fe (iron) in aerosol collected above the sea as a target for demonstration. 

Iron is abundant in the Earth since it is the most stable nucleus of all the elements. It also plays an important role in controlling the Earth's environment. For example, the amount of Fe at the surface of the ocean affects the number of phytoplankton, which is related to the uptake of CO$_2$, and thus has implications for global climate change\cite{1988Natur.331..341M}\cite{Moore2013}. One of the important sources of Fe at the ocean surface is aerosols, containing various species of Fe that originate from natural and anthropogenic sources\cite{Jickells67}\cite{doi:10.1246/cl.160451}. The Fe species is important since highly soluble Fe species are preferentially dissolved in seawater and utilized by phytoplankton\cite{doi:10.1098/rsta.2016.0190}\cite{SHOLKOVITZ20093981}. Although XAFS is an effective method 
to measure the speciation of iron\cite{Schroth2009}\cite{acp-13-7695-2013}, precise measurement of marine aerosols is challenging since the amount of the aerosols above the sea is 
about three orders of magnitude lower than that collected on land.

The aerosol specimen was prepared in the same way as described in \cite{kurisu2019}. 
The aerosol sample was collected during the $R/V$ Hakuho-maru KH-17-3 cruise in the subarctic North Pacific. 
The aerosols were separated into six size-fractions, and one of them (size fraction: 0.49--0.95 $\mu$m) was used for the analysis. 
Since the amount of the collected samples is very small, an acid-washed polytetrafluoroethylene (PTFE) sheet (Naflon tape; thickness = 0.2 mm; Nichias Co., Ltd., Japan) was used to minimize contamination from the sampling filter. Details on the preparation methods of the PTFE filter are described in \cite{SAKATA2018100}. 
The amount of Fe of the sample was several ng of Fe per 1cm$^2$ filter, 
which is measured by using inductively coupled plasma-mass spectrometry (ICP-MS, Agilent 7700) after acid digestion of the aerosol sample 
based on our previous work\cite{SAKATA2018100}. 
Since we could not eliminate the contamination from Fe in the filter or the beamline, 
we took data from the aerosol sample on the PTFE filter, 
and from a PTFE filter without an aerosol sample (a blank target).

The input energy ranges were 7086.9 -- 7180.4 eV for the aerosol and 7087.2 -- 7165.2 eV for the blank target 
to sweep from the pre-edge energy of Fe-K to the post-edge energy. 
The TES count rate was adjusted to be around 2~kc/s/ array by adjusting the slit when the beam energy is set at the post-edge region. 
Figure \ref{fespec} shows the XES of the aerosol sample and the blank target, 
which are created by using all the time during the sweep just to clarify the difference in the spectra. 
The two XES are normalized by integrating the counts/$I_0$, 
where both are a function of the beam energy. 
The excess of the weak lines from Ca, V, Cr, Mn, and Fe come from the sample. 
The width of the beam energy sweep is 93.3~eV (=7180.4-7086.9~eV) for the aerosol, 
and the Bi escape peaks seen at 4600-4700 eV are broad due to the sweep of the input energy. 
The same feature is seen in the XES of the blank target. 
For a reference, the XES of SDD taken at a beam energy of 7131.58 eV is shown in figure \ref{fespec}. 
The interference of scattered X-rays at Fe region is more severe in SDD. 

\begin{figure}[htbp]
\begin{center}
\includegraphics[width=0.95\linewidth, keepaspectratio]{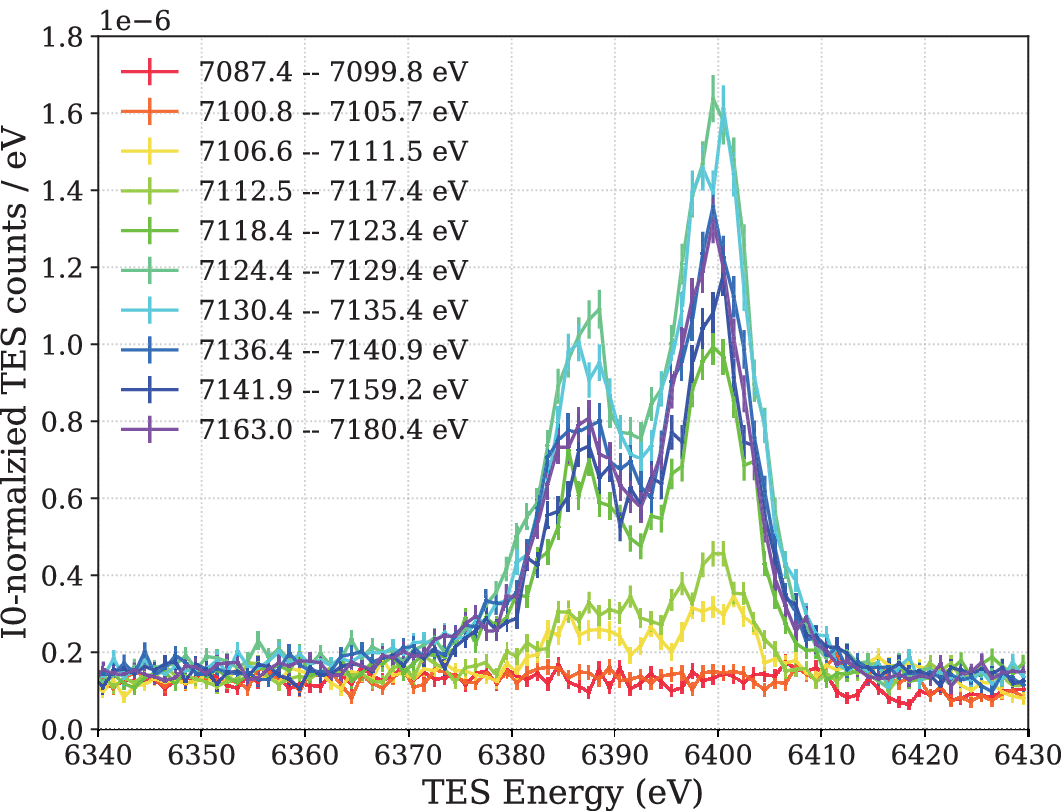}
\includegraphics[width=0.95\linewidth, keepaspectratio]{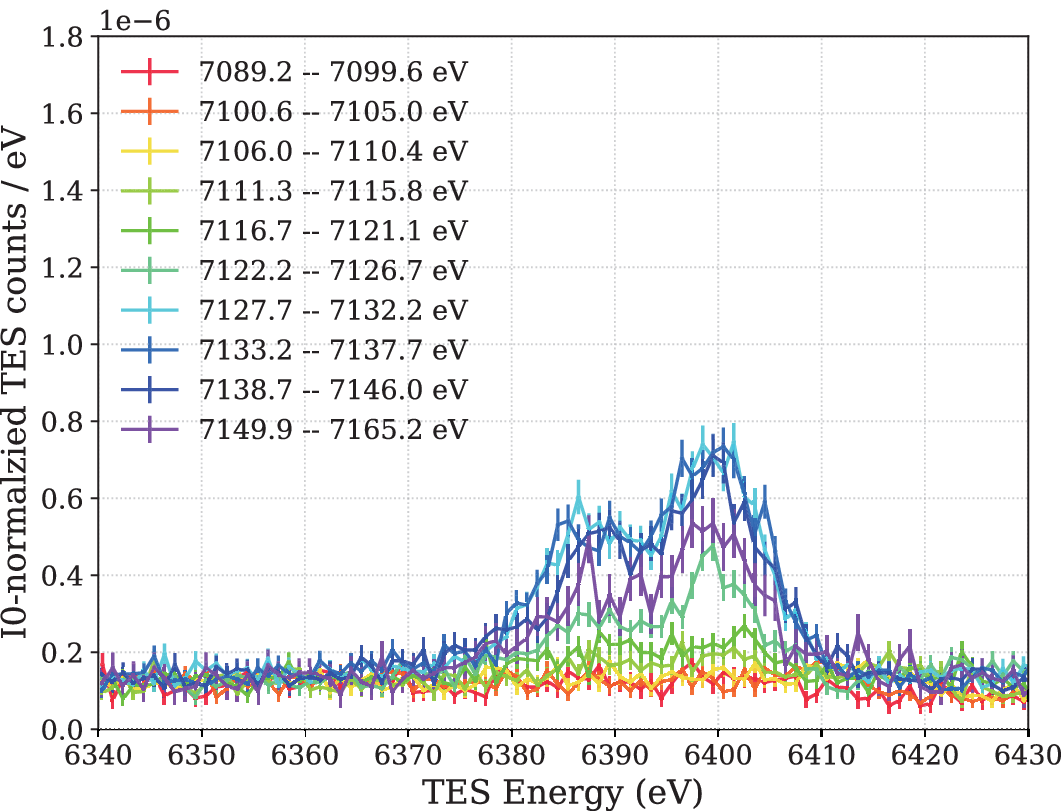}
\caption{(top) The Fe K$\alpha1$ and K$\alpha2$ spectra of the aerosol specimen for ten different beam energies, which are specified in the legend by different colors. 
The error bars refer to the statistical errors on the number of photons per energy bin. (bottom) The same measurement as above, repeated for the blank target.}
\label{fe_spec_comp}
\end{center}
\end{figure}

Figure \ref{fe_spec_comp} shows the XES of the two targets as the beam energy was swept across the Fe K-edge energy. 
The entire range of the scan is divided into ten segments and shown in different colors in figure \ref{fe_spec_comp}. 
For reference, the count rate of the TES in the 6330--6430 eV range for the aerosol specimen was 455c/60s when the beam energy was 7086.9 keV and 1327c/60s when the beam energy was 7180.4 eV. 
For the blank target, count rates were 183c/30s at 7087.2 keV beam energy and 308c/30s at 7165.2 eV and $I_0$ did not significantly change during the sweep, which means the counts from background Fe K lines were 25~c/30s (=308-183 c/30s) . The difference on the integration time (60s for the sample, 30 s for the blank) is due to a schedule constraint. 
The energy resolution of the spectra in figure \ref{fe_spec_comp} is estimated from a valley between Fe K${\alpha 1}$ and Fe K${\alpha 2}$. 
A similar approach is used with manganese (see figure 7\cite{10.1093/pasj/psz053}). 
This is a practical way of estimating the energy resolution, because calibration uncertainties and cross talk, and operational environments of the TES could cause the failure of the fitting due to unexpected tails around the peaks. 
For the current measurement, a typical statistical error in each spectral bin is larger than a few percent, which means that it is difficult to assess chemical shifts of a peak energy, typically below 1~eV \cite{2005NiK}. Also, another desirable analysis is to obtain the information on resonant inelastic X-ray scattering (RIXS)\cite{2008Hahashi}, 
though the statistical errors due to high background and a limited number of photons hinder further analysis. Larger effective area with more pixels and lower background would enable further improvement in application.

\begin{figure}[htbp]
\begin{center}
\includegraphics[width=0.95\linewidth, keepaspectratio]{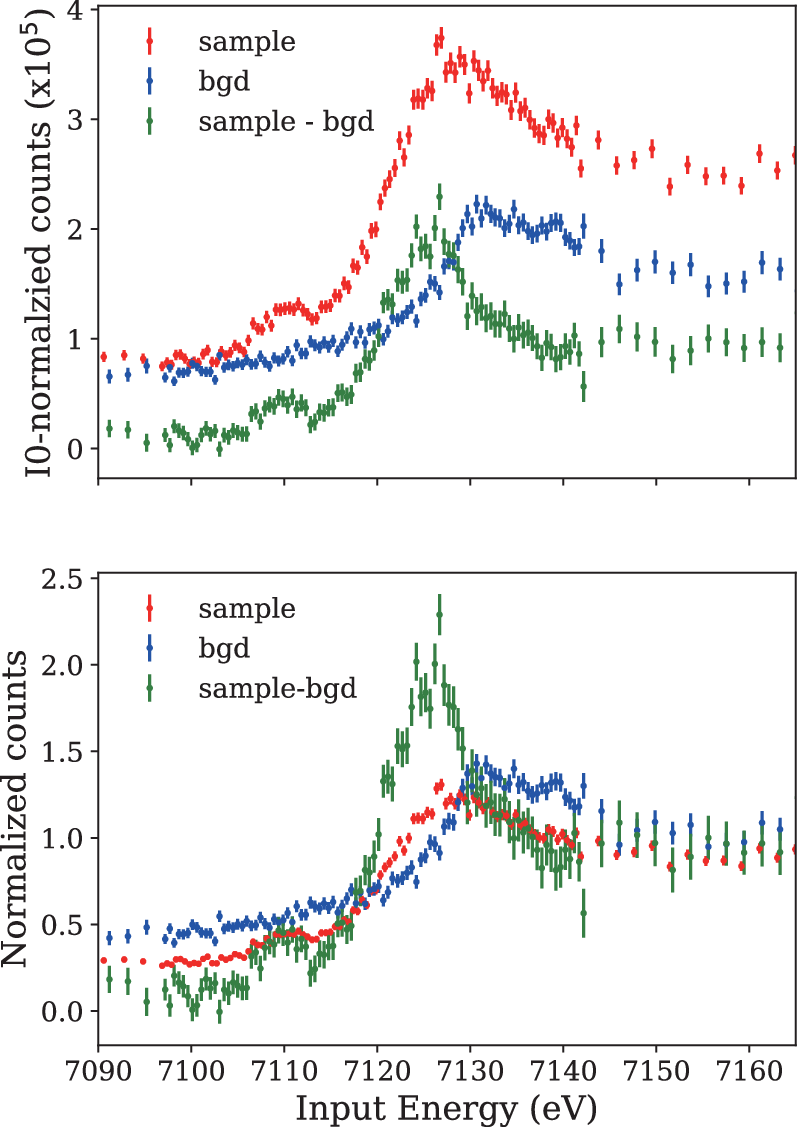}
% /Users/syamada/work/ana/from_old_mac/HEATES/SPring-8_201907/Fe_Kurisu/sp8_comp_Fe_paper_withoutsdd.py
\caption{(top) XANES spectra of the aerosol sample, the blank target, and the difference of the two, are shown in red, blue, and green, respectively. The horizontal axis is the X-ray energy of the beam, while the vertical axis is the TES counts in the 6360--6420 eV range normalized by $I_0$. (bottom) The same as above, with the spectra normalized to their mean values at the higher end of the energy range. } 
\label{fexanes}
\end{center}
\end{figure}

The XANES spectra are created by plotting the sums of the counts in TES from 6360 eV to 6420 eV 
and normalizing by $I_0$. 
The difference of the dead time fraction during the scan is corrected by using the good and bad event fractions, 
which are almost constant at $\sim$ 3\% across the input energy. 
Figure \ref{fexanes} shows the XANES spectra of the aerosol specimen, the background, and the difference of the two. 
The subtraction is performed in the unit of counts per $I_0$. This is based on the assumption that the background is scaled by $I_0$, which is valid assuming the spatial distributions of the scattered, absorbed, and transmitted photons do not change over the specimen.  
The assumption is only valid when the specimen is thinner than the total attenuation length. 
However, if the specimen is thicker than the total attenuation length, 
it could change the irradiation environment and thus the background estimation using the blank target would be more difficult. 
The aerosol sample is almost invisible to the unaided eye, so it is in the former case. 

%The XANES spectrum obtained with the SDD is overlaid in figure \ref{fexanes}. The TES spectra showed a better signal to background ratio (the ratio of $I_0$-normalized counts at the peak and pre-edge energies; $\sim$ 5.2) compared with the SDD spectra ($\sim$ 1.1) because the energy resolution of the TES reduces the interference of scattered X-rays and contamination from background components. The difference between the TES and SDD is qualitatively significant.  %The peak and shape obtained with SDD remain entangled with the contamination of the background components. 

\begin{figure}[htbp]
\begin{center}
\includegraphics[width=0.95\linewidth, keepaspectratio]{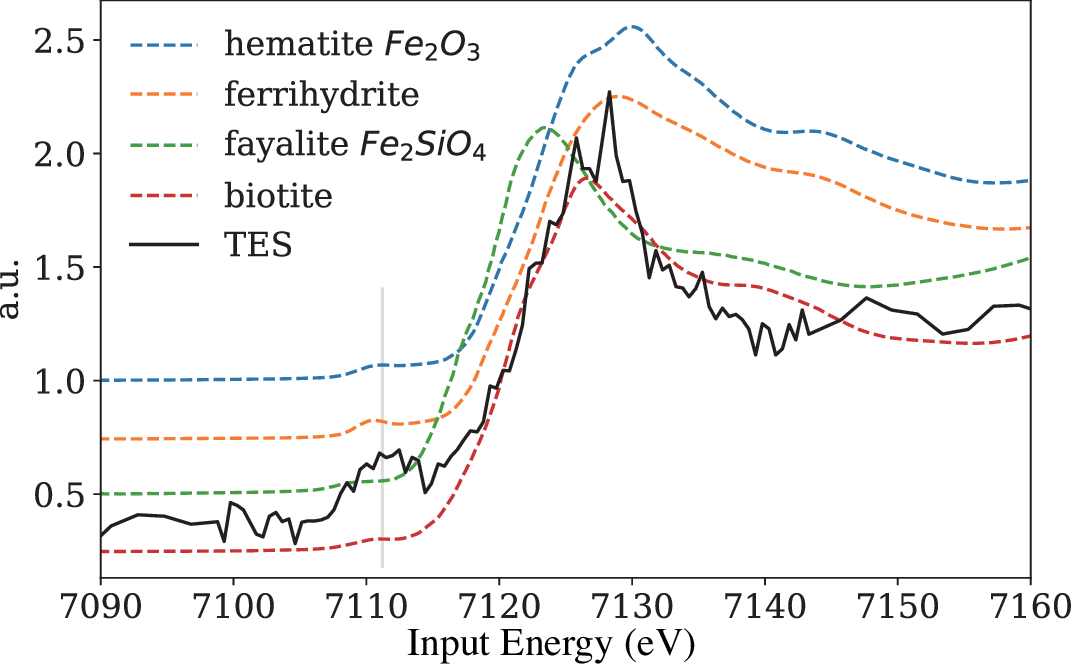}
%/Users/syamada/work/ana/from_old_mac/HEATES/SPring-8_201907/Fe_Kurisu/plot_comp_Fe_XANES_forPapar/for-revise_energy-shifted_/sp8_paper_comp_fe_xanes_shifetd.py
\caption{
Fe K-edge XANES spectra of reference compounds and the aerosol sample measured by the TES; hematite (blue), ferrihydrite (orange), fayalite (green), biotite (red), and the TES spectrum (black).  The vertical axis is a normalized XANES spectra with constant offsets. A vertical line refers to 7111.2 eV, corresponding to 1s to 3d/4p transition of hematite.}
\label{fexanes_model}
\end{center}
\end{figure}

The XANES of TES is compared to several possible references in figure \ref{fexanes_model}, 
including XANES data of hematite, ferrihydrite, fayalite, and biotite that we obtained by ourselves in transmission mode at the PF BL-12C in KEK. 
Hematite measured both at KEK and SPring-8 is used for calibration between the two beam lines. 
The energy of a pre-edge peak, corresponding to 1s to 3d/4p transition of hematite, 
is 7111.2 eV as indicated by a vertical line in figure \ref{fexanes_model}. This is used as an energy calibration of the beam energy. 
Of the reference spectra, the TES spectrum seems closest to that of biotite. 
It is confirmed by conducting the least-square fitting of the XANES spectra 
and the spectrum measured by TES is best fitted by that of biotite. 
Biotite is abundant in mineral dust aerosols with low Fe solubility, 
which could be a reasonable scenario for the source of this sample. 
By reducing contamination of Fe from the filter or the beamline and by estimating the background as above, 
the TES can be a powerful tool to investigate dilute quantities of Fe species in aerosols. 

\section{\label{sec:level3}Summary and discussion}

% general summary
The performance of a TES-based spectrometer for high energy resolution X-ray spectroscopy has been successfully 
demonstrated at the third generation synchrotron facility SPring-8. 
The TES spectrometer used was a NIST 240 pixel TES, which is a stable, technically mature instrument, and has therefore been deployed for many experiments worldwide. The utility of the TES for a particular XAFS application needs to be determined by considering the requirements for the energy resolution and the photon statistics in the region of the interest. These parameters are not solely determined by the design of the TES spectrometer; 
the integration of the TES into the beamline and the sample under study also affect the count rates measured by the TES, 
and consequently affect the energy resolution. The accuracy of the energy scale, which is necessary to measure the chemical shifts, depends on the availability of calibration lines in the scientific region of interest. Calibration lines must be produced with sufficient photon statistics for an accurate measurement, and effort in post-processing is required to meet specific calibration requirements. 
Therefore, successful application of a TES spectrometer at a synchrotron facility involves a series of collaborative and continuous efforts to optimize the performance of the detector and the accelerator to accomplish a particular scientific goal.

\begin{figure}[htbp]
\begin{center}
\includegraphics[width=0.99\linewidth, keepaspectratio]{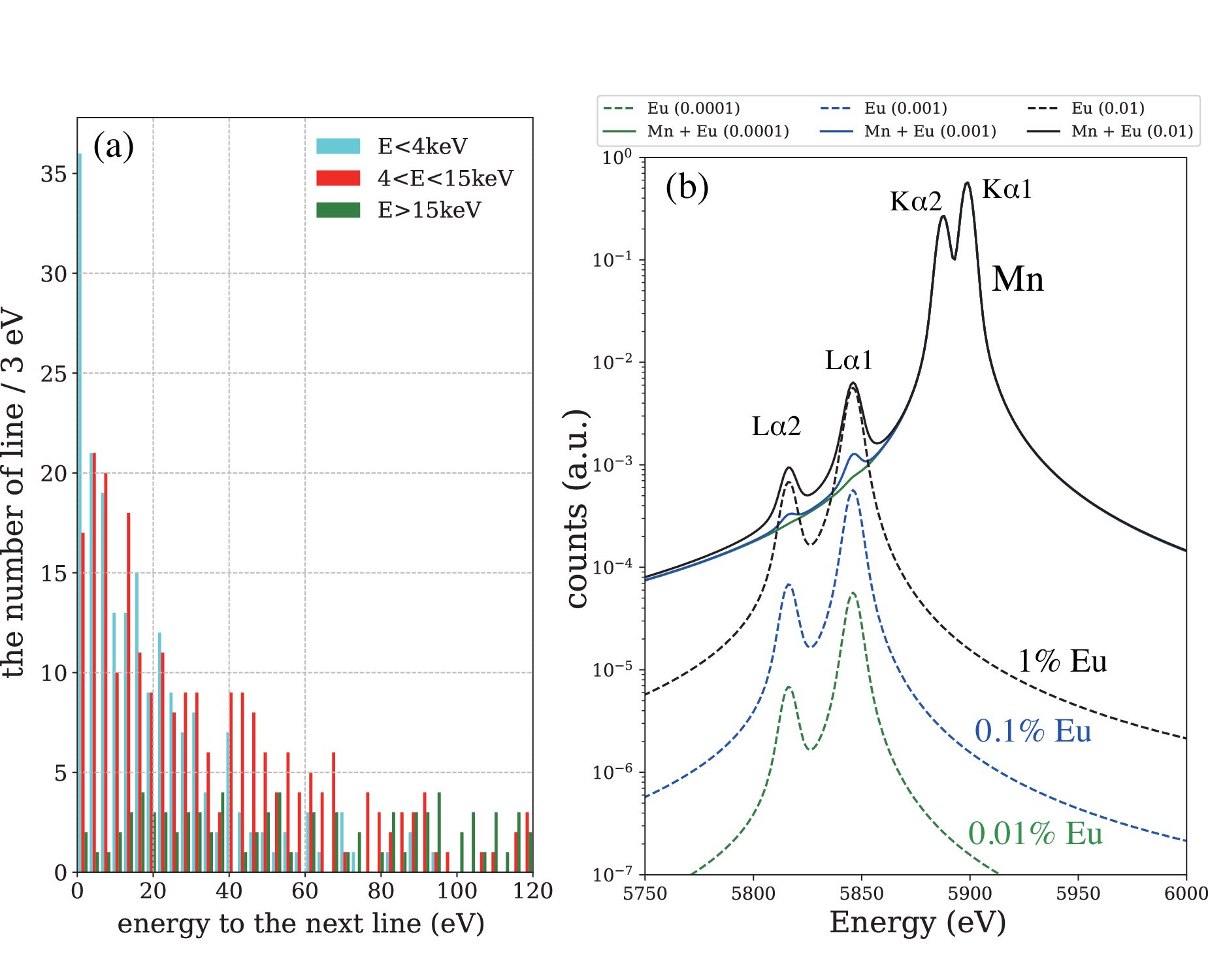}
%\includegraphics[width=0.95\linewidth, keepaspectratio]{figures/x-ray_booklet/xray_booklet_ene.eps}
%/Users/syamada/work/ana/from_old_mac/HEATES/SPring-8_201907/x-ray_booklet/process_xray_booklet_forpaper.py
%/Users/syamada/work/ana/sp8tes/calc_calsource/20191214_calibration/Programs/forHayakawakun20200318/xray_tools_forpaper.py
\caption{(a) Distribution of the energy differences between one line and the next. 
The vertical scale is the number of fluorescence lines per 3 eV bin.
The blue, red, and green ones refer to the samples in $E<4$ keV, $15>E>4$ keV, and $E>15$ keV, respectively. 
(b) Simulated spectra of Mn and Eu using natural widths given an energy resolution of 5~eV. The efficiency shown in figure \ref{tesqe} is included, though it does not change the 
relative intensity within the narrow energy range. The mass fraction of Mn to Eu is 1:0.01 (black), 1:0.001 (blue), and 1:0.0001 (green). 
}
\label{linehist}
\end{center}
\end{figure}

% TES specific development and summary to be noticed 
The performance of the TES spectrometer was studied over the energy range 4--13 keV, using a Mn target with varying count rates and a Pb target with varying beam energies. The energy resolution is approximately 5~eV FWHM at 6~keV at  total count rates of $<$ 2~kc/s/array, 
which increases to a few tens of eV above 10~keV. %The dead time below a count rate of $\sim$ 2~k c/s/array was less than 5 percent. 
Although there is still room for improvement in the energy resolution, 
we have demonstrated the ability of TES to simultaneously detect multiple elements in a standard sample. 
There are many lines that can be successfully separated by the TES, even with the current energy resolution. 
Figure \ref{linehist} (a) summarizes the number of emission lines as a function of the energy gap between lines, 
using data from the X-ray Data Booklet compiled by Lawrence Berkeley National Laboratory\cite{thompson2001x}.
%all 700 above = 459 % = 65.5714285714
%loew 198 above = 74 % = 37.3737373737
%highe 257 above = 155 % = 60.3112840467
%high2e 243 above = 228 % = 93.8271604938
Only the strongest lines for atomic numbers 3 $ \leqq Z \leqq $ 95 are included:
K$\alpha1$, K$\alpha2$, K$\beta1$, L$\alpha1$, L$\alpha2$, L$\beta1$, L$\beta2$, L$\gamma1$, and M$\alpha1$, which results in a total of 700 emission lines. 
The lines are divided into three energy ranges: $<$ 4 keV, 4--15 keV, and $>$ 15 keV. 
The number of pairs of lines with an energy difference larger than 20 eV is 74 out of 198 pairs of lines in $E <$ 4 keV, 155 out of 257 pairs in 4 $< E < $ 15 keV, and 228 out of 243 pairs in E $>$ 15 keV, respectively. 
As an example of a realistic application, expected spectra of Mn and Eu are shown in figure \ref{linehist} (b). 
This assumes a study on rare earth elements in a natural sample. 
The efficiency shown in figure \ref{tesqe} is included in the calculation, and the lines with natural widths 
are convolved with an energy resolution of 5 eV without a tail component.  
The mass fraction of Eu has changed from 0.01\%, 0.1\%, and 1\% of Mn. 
%Without calculating a statistical error, in the lowest case the detection of Eu is difficult due to the tail of the natural width of Mn K lines.  
The detection of Eu could be possible when its mass fraction is larger than 0.1\% of Mn, 
though it could be more difficult as it is smaller. 
Therefore, although there are many lines that can be distinguished by the energy resolution of the TES in the hard X-ray energy range, 
the requirements on experiments must be carefully assessed for each sample.  
One must consider the energy resolution, photon statistics, accuracy of the energy calibration, and the relative intensities of different atoms around the region of interest. 

% XAFS related summary 
We have also demonstrated a measurement of XANES from Fe in a dilute environmental sample. 
In a practical application, Fe is one of the most difficult elements for the measurement 
because Fe is present in many of the support structures in both the beamline and the TES system. 
As shown in figure \ref{fespec}, 
the continuum of the background spectrum is almost the same as that of the sample. 
In this particular case, the difference between the spectrum of the sample and that of the blank target, 
which are normalized by $I_0$, can be identified as the excess of the weak lines from Ca, V, Cr, Mn, and Fe in the sample.
Even if the value of $I_0$ is not accurately known, the normalization factor can be estimated 
by comparing the background continuum level. %since their shape is almost the same. 
The method is only valid when the target is very thin in terms of the total attenuation length; 
e.g.,  $1/\sigma_{Fe}(6~\rm{keV}) \sim 0.012$[g/cm$^2$] and $1/\sigma_{Fe}(14~\rm{keV}) \sim 0.014$[g/cm$^2$]. 
When the target is thick, it should change the shape of the background continuum and the relative intensities of the emission lines.
In such a case, the absolute value of $I_0$ can be a unique reference for the comparison. 
However, setting the same level of $I_0$ for both the sample and the blank is operationally difficult. 
If the input X-ray rate at the absorber of the TES when a beam energy is in pre-edge region 
is adjusted to be the same for both the sample measurement and the blank,  
the time for the measuring the blank could be longer than that when the beam intensity is optimized. 
In other words, once the setup is optimized without the sample, the input rate for the thick sample would be too large. 
Furthermore, if background lines originate somewhere along the beamline, such as in the pipes of the beamline that are made from alloys of Fe or chromium, 
the assumption that the background is scaled by $I_0$ may not hold. 
Although more systematic studies are needed to assess methods of background subtraction for thick samples, 
it is still the case that identifying the origin of the background components is aided by the high energy resolution of the TES, 
resulting in a better understanding of the environment of the experimental setup. 

% future prospect 
Development of TES technology is ongoing, and will lead to performance improvements in terms of the number of pixels, the effective area, and the energy range of the sensor arrays. 
One of the breakthrough methods that has seen rapid progress in recent years is the readout of TES signals via microwave SQUID multiplexing\cite{doi:10.1063/1.4986222}\cite{Yoon2018}, 
which enables us to read about hundreds of pixels with only a few coaxial cables required to run from the cold readout stage to the room temperature electronics. 
Microwave multiplexing also provides a significant increase in available readout bandwidth, enabling the readout of larger arrays and faster signals. 
Real-time processing capabilities are also under development, which are needed for the TES to be widely accepted by the X-ray science community.
%When TES instruments have larger arrays, they will be more 
%When the number of pixels is more significant, 
%the optimal solution for the implementation is not unique and expensive in resources of both hardware and software. 
Future space applications or other remote instruments will require practical solutions; e.g., a space-flight compatible electronics \cite{10.1117/1.JATIS.5.2.021013} 
in terms of the detailed assessment on the digital electronics. 
When new technological developments enable more TES pixels, and the energy resolution approaches the practical limit ($\sim$eV),  
the TES will find new applications in the study of spatial and steric microstructure, and for high-resolution XANES analyses such as RIXS\cite{2008Hahashi} and the measurement of chemical effects\cite{BOYDAS20151757}. Our results are useful as a benchmark of the performance of the current generation of TES spectrometers at a hard X-ray synchrotron facility. 

\begin{acknowledgments}

This work was partly supported by the Grants-in-Aid for Scientific Research (KAKENHI) from MEXT and JSPS (No. 16H02190, 18H05458, 18H03714, 18H01260, 18H03713, 18H03892, 17H06455, 20K20527 and 19K21884), and RIKEN Pioneering Project: Evolution of Matter in the Universe. 
The authors are grateful to the members in the NIST Quantum Sensors Project. 
We appreciate the significant contributions by SPring-8, J-PARC, RIKEN, and 
those who have backed up SPring-8  2019A1523, 2019B1498, and 2020A0174 experiments. 
\end{acknowledgments}

\section*{Data Availability Statement}
The data that support the findings of this study are available from the corresponding author upon reasonable request.

 %\nocite{*}
\bibliographystyle{unsrt}
\bibliography{yamadates}% Produces the bibliography via BibTeX.

\end{document}